\documentclass[12pt]{article}
\usepackage{amsmath,amssymb,amsthm,amsxtra,overpic,bbm,bm,epsfig,subfigure}
\usepackage{hyperref}
\usepackage{mathrsfs}
\usepackage{graphicx}
\usepackage{multirow}
\usepackage{color}
\usepackage{comment}
\usepackage{epstopdf}
\usepackage{float}
\usepackage{cite}
\usepackage{hyperref}
\usepackage{url}
\usepackage{slashed,stmaryrd}

\def\thefootnote{\fnsymbol{footnote}}

\begin{document}
\vspace{0.2cm}
\begin{center}
{\Large\bf General Remarks on the One-loop Contributions to \\ the Muon Anomalous Magnetic Moment}
\end{center}
\vspace{0.2cm}

\begin{center}
{\bf Bingrong Yu}~$^{a,~b}$~\footnote{E-mail: yubr@ihep.ac.cn},
\quad
{\bf Shun Zhou}~$^{a,~b}$~\footnote{E-mail: zhoush@ihep.ac.cn (corresponding author)}
\\
\vspace{0.2cm}
{\small
$^a$Institute of High Energy Physics, Chinese Academy of Sciences, Beijing 100049, China\\
$^b$School of Physical Sciences, University of Chinese Academy of Sciences, Beijing 100049, China}
\end{center}

\vspace{0.5cm}

\begin{abstract}
The latest measurement of the muon anomalous magnetic moment $a^{}_{\mu} \equiv (g^{}_\mu - 2)/2$ at the Fermi Laboratory has found a $4.2\,\sigma$ discrepancy with the theoretical prediction of the Standard Model (SM). Motivated by this exciting progress, we investigate in the present paper the general one-loop contributions to $a^{}_\mu$ within the SM and beyond. First, different from previous works, the analytical formulae of relevant loop functions after integration are now derived and put into compact forms with the help of the Passarino-Veltman functions. Second, given the interactions of muon with new particles running in the loop, we clarify when the one-loop contribution to $a^{}_\mu$ could take the correct positive sign as desired. Third, possible divergences in the zero- and infinite-mass limits are examined, and the absence of any divergences in the calculations leads to some consistency conditions for the construction of ultraviolet complete models. Applications of our general formulae to specific models, such as the SM, seesaw models, $Z^\prime$ and leptoquark models, are also discussed.
\end{abstract}

\newpage

\def\thefootnote{\arabic{footnote}}
\setcounter{footnote}{0}

\section{Introduction}
The Standard Model (SM) of particle physics, based on the relativistic quantum field theory with gauge symmetry ${\rm SU}(3)^{}_{\rm C} \otimes {\rm SU}(2)^{}_{\rm L} \otimes {\rm U}(1)_{\rm Y}^{}$, has proved to be a very successful theory in describing strong, weak and electromagnetic interactions among all discovered elementary particles. Recently, the Muon $(g - 2)$ Collaboration at Fermi National Accelerator Laboratory has announced their precise measurement of the muon anomalous magnetic moment $a_{\mu}^{} \equiv \left(g_{\mu}^{}-2\right)/2$ with $g_{\mu}^{}$ being the spin $g$-factor of muon~\cite{Fermilab2021}, which combined with the previous E821 measurement at Brookhaven National Laboratory in 2006~\cite{Bennett2006} leads to a $4.2\,\sigma$ discrepancy with the SM prediction~\cite{Aoyama2020,Davier2010,Davier2017,Davier2019}, namely,
\begin{eqnarray}
\Delta a_{\mu}^{}\equiv a_{\mu}^{\rm Exp}-a_{\mu}^{\rm SM}=\left(251\pm 59\right)\times 10_{}^{-11} \; .
%     (1)
\label{eq:Deltaamu}
\end{eqnarray}
If confirmed by more precise measurements and theoretical predictions~\cite{Aoyama2020, Borsanyi2020}, such a discrepancy will definitely signify new physics beyond the SM (BSM). There have already been a great number of theoretical works attempting to explain this discrepancy by new physics~\cite{Zhang2021,Arcadi2021DM,Zhu2021,Nomura2021,Baum2021,Endo2021,Ge2021,
Ahmed2021,Das2021,Han2021,Keung2021,Bai2021,Brdar2021,Zu2021,Ibe2021,
Cox2021,Abdughani2021,VanBeekveld2021,Buen-Abad2021,ChenChuanHung2021,Babu2021,Ferreira2021,Yin2021,
Hanchengcheng2021,Gu2021,Wangfei2021,Calibbi2021,Wang2021,Litianjun2021,
Cadeddu2021,Chen2021,Escribano2021,Chun2021,Aboubrahim2021SUSY,
Bhattacharya2021,Yang2021,Arcadi2021,Lu2021,Chakraborti2021GUT,LiTong2021,
Cen2021,Borah2021,Marzocca2021,Du2021,Zhou2021,Ban2021,Anchordoqui2021,
CarcamoHernandez2021,Baer2021,Altmannshofer2021,Balkin2021,
Cacciapaglia2021,Dasgupta2021,Aboubrahim2021,Ma2021,Jueid2021,
FileviezPerez2021,Baryshevsky2021,Ghorbani2021,Carpio2021,Arbuzov2021,
Alvarado2021,Chakraborti2021,Chang2021,Zheng2021,Dutta2021,Zhangdi2021,
Hou2021,Cirigliano2021,Allwicher2021,Jia2021,De2021,Dey2021,Arkani-Hamed2021, Jeong2021,Li2021,Crivellin2021,Cao2021}. See, e.g., Refs.~\cite{Jegerlehner2009,Lindner2016,Athron2021,Keshavarzi2021}, for comprehensive reviews on this topic and also for more earlier references.

In order to account for the deviation $\Delta a^{}_\mu$ in Eq.~(\ref{eq:Deltaamu}) by new physics, one may just introduce the BSM particles and their interactions with the SM particles. In comparison with a specific model, such a phenomenological approach is less model-dependent and has been taken up in many previous works. As a matter of fact, the one-loop contributions to $\Delta a^{}_\mu$ in this approach have been systematically calculated in Ref.~\cite{Leveille1977} and later verified in Ref.~\cite{Moore1984}. In these two papers, the final results of $\Delta a^{}_\mu$ have been given in terms of uncompleted integrals, since the Feynman parametrization has been used to deal with the one-loop calculations and the integration over the Feynman parameter is left intact. The approximate expressions of these one-loop contributions at the leading order of small particle-mass ratios calculated in the Feynman gauge are collected in Ref.~\cite{Athron2021}, as well as in some earlier literature ~\cite{Queiroz2014,Crivellin2018}.

Two shortcomings of the existing results in Refs.~\cite{Leveille1977,Moore1984} should be noticed. First, for a general interaction without assuming hierarchical particle masses, it is usually difficult to integrate the Feynman parameter out and derive the final analytical results. Second, even with some assumptions of particle masses, one must be extremely careful with the order of series expansions and the integration. For instance, the contribution to $a^{}_\mu$ from the Higgs boson in the SM is given by~\cite{Leveille1977}
\begin{eqnarray}
\label{eq:integrand}
a_{\mu}^{{\rm SM\,Higgs}}\propto \int_0^1 \frac{x_{}^2\left(2-x\right)}{1-x+x^2_{}\left(m_\mu^2/ M_H^2 \right)} {\rm d}x \;,
%     (2)
\end{eqnarray}
where $m_{\mu}^{} \approx 106~{\rm MeV}$ is the muon mass and $M_H^{} \approx 125~{\rm GeV}$ is the Higgs-boson mass. Although $m_{\mu}^{}\ll M_H^{}$ is perfectly satisfied, it is problematic to first expand the integrand in Eq.~(\ref{eq:integrand}) as a series of the small mass ratio $m_{\mu}/M_H^{}$ and then integrate them out order by order. The reason is simply that the integral at each order is actually divergent, whereas the whole integral in Eq.~(\ref{eq:integrand}) is finite as it should be.

In this work, we are thus motivated to revisit the one-loop contributions to $a^{}_\mu$ and derive the final analytical formulae. First, instead of the Feynman parametrization, we implement the Passarino-Veltman functions~\cite{Passarino1978} and present the results in compact forms by using the publicly available Package-X~\cite{Patel2015,Patel2017}. Without any integrals in the final formulae, one can safely expand those general expressions as the series of the small mass ratios without any divergences, which should be helpful for phenomenological studies. Second, while there have been many attempts in the literature to find positive contributions to $\Delta a_{\mu}^{}$ (e.g., the two-Higgs-doublet model and the $Z_{}^{\prime}$ model), some contributions emerge with a wrong sign (e.g., type-I and type-II seesaw models for neutrino masses). One immediate question is whether one can set up a simple criterion for positive contributions to $\Delta a^{}_\mu$ in an arbitrary new-physics model without repeating tedious loop calculations. Given all the interactions between the BSM particles and the SM particles, as well as the basic properties of those particles (e.g., masses, spins and electric charges), we find that with reasonable assumptions of particle masses it is indeed possible to make a quick judgment about the sign of $\Delta a_{\mu}^{}$. For example, this goal can be achieved according to some simple criteria on the model parameters, such as the electric charges of the particles running in the loop and the relative sizes of the coupling constants.

The remaining part of the present paper is organized as follows. In Sec.~\ref{sec:loop}, following the earlier works~\cite{Leveille1977, Moore1984}, we classify the one-loop contributions to $a^{}_\mu$ by the particles running in the loop. More explicitly, there are three different categories: (1) a fermion and a scalar boson (i.e., the FS-type); (2) a fermion and a vector boson (i.e., the FV-type); and (3) a fermion, a scalar boson and a vector boson (i.e., the FSV-type). For each type of models, general expressions of $a^{}_\mu$ without any approximations are derived by using the Passarino-Veltman functions. Some simplifications of the formulae in various limits of hierarchical particle masses are discussed. The general formulae derived in Sec.~\ref{sec:loop} are then applied in Sec.~\ref{sec:examples} to some concrete models that have been extensively discussed in the literature. We also demonstrate how to make a quick judgment about the sign of $\Delta a_{\mu}^{}$ for a concrete model. Finally, our main conclusions are summarized in Sec.~\ref{sec:summary}.

\section{One-loop Contributions}
\label{sec:loop}

Since we are interested in the extra one-loop contributions to $a^{}_\mu$ from the BSM particles, it is necessary to specify the particles running in the loop, given the external particles (i.e., the muon and the photon). Without resorting to a specific model, one can follow the phenomenological approach in Refs.~\cite{Leveille1977, Moore1984} and write down the interaction terms for the BSM and SM particles. As already mentioned above, there are three possible scenarios in which the basic symmetries, such as the Lorentz invariance and the ${\rm U}(1)$ gauge symmetry of quantum electrodynamics (QED), are required to be preserved.

\subsection{The FS-type}

\begin{figure}[t!]
	\centering		\includegraphics[width=17.5cm]{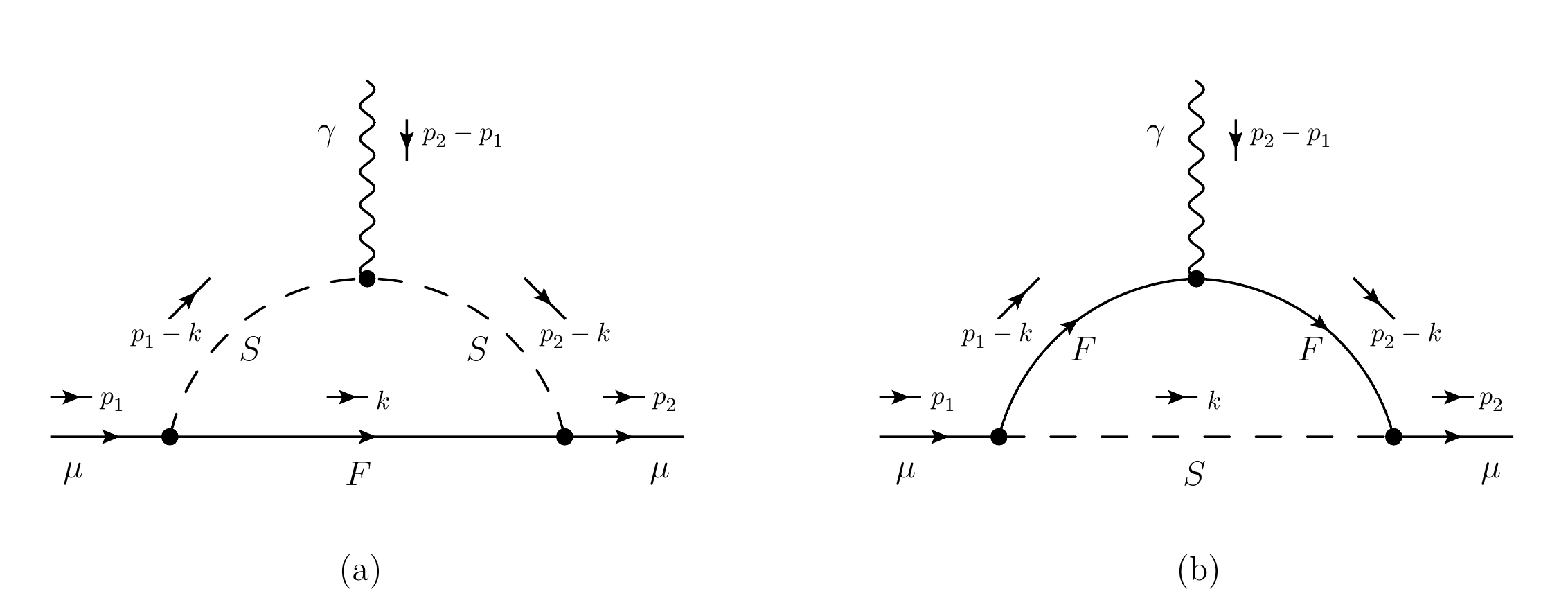}
	\vspace{-0.8cm}
	\caption{The Feynman diagrams for the FS-type contributions to the muon anomalous magnetic moment $a^{}_\mu$, where the particle four-momenta have been explicitly specified and represented by arrows along the internal and external lines.}
	\label{fig:FS}
\end{figure}

First, we consider the case where there is a fermion $F$ and a scalar boson $S$ in the loops that contribute to $a^{}_\mu$. The Feynman diagrams are shown in Fig.~\ref{fig:FS}. These FS-type diagrams appear in many well-known models, such as the SM Higgs, the two-Higgs-doublet model, the type-II seesaw model, and the scalar leptoquark model.

After the spontaneous symmetry breaking of ${\rm SU}(2)^{}_{\rm L}\otimes {\rm U}(1)^{}_{\rm Y}$ to ${\rm U}(1)^{}_{\rm QED}$, the interactions among the newly-introduced particles and the relevant SM particles can be written as
\begin{eqnarray}
\label{eq:FS langrangian}
{\cal L}_{\rm int}^{\rm FS}=\left[\overline{F}\left( C_{\rm S}^{}+ C_{\rm P}^{}\gamma_5^{}\right)\mu S_{}^{\dagger}+{\rm h.c.}\right]+Q_F^{}e\overline{F}\slashed{A}F+{\rm i}Q_S^{}e\left(S_{}^{\dagger}\partial_{\mu}^{}S-S\partial_{\mu}^{}S_{}^{\dagger}\right)A_{}^{\mu}\;,
%     (3)
\end{eqnarray}
where $C_{\rm S}^{}$ and $C_{\rm P}^{}$ stand respectively for the scalar-type and pseudo-scalar-type coupling constants. The interaction between $S$ and the photon is determined by the minimal coupling arising from the kinetic term of $S$ via the replacement $\partial_{\mu}^{}\rightarrow\partial_{\mu}^{}-{\rm i}Q_S^{}eA_{\mu}^{}$ as in the ordinary QED, so is the interaction between the fermion $F$ and the photon, where $Q^{}_S$ and $Q^{}_F$ denote the corresponding electric charges in units of the electric charge of positrons. Due to the conservation of electric charges, we always have $Q^{}_F + Q^{}_S = -1$, which will be implied in the subsequent discussions in this subsection.
%%%%%%%%%%%%%%%%%%%%%%%%%%%%%%% Fig. 1 %%%%%%%%%%%%%%%%%%%%%%%%%%%%%%%%
%%%%%%%%%%%%%%%%%%%%%%%%%%%%%%%%%%%%%%%%%%%%%%%%%%%%%%%%%%%%%%%%%%%%%%%

Now it is straightforward to find out the amplitudes corresponding to the Feynman diagrams in Fig.~\ref{fig:FS}(a) and Fig.~\ref{fig:FS}(b), namely,
\begin{eqnarray}
\label{eq:FSa amp}
\overline{u}(p_2^{})\left(-{\rm i} e \Gamma_{{\rm FS}}^{\mu,({\rm a})}\right)u(p_1^{})
&=&{\rm i} Q_S^{}e \int \frac{{\rm d}^4 k}{\left(2\pi\right)^4}\left[\overline{u}(p_2^{}){\rm i}\left(C_{\rm S}^{*}-C_{\rm P}^{*}\gamma_5^{}\right)\frac{{\rm i}}{\slashed{k}-M_F^{}}{\rm i}\left(C_{\rm S}^{}+C_{\rm P}^{}\gamma_5^{}\right)u(p_1^{})
\right.\nonumber\\
&&\left. \times \frac{{\rm i}}{\left(p_2-k\right)^2-M_S^2}\frac{{\rm i}}{\left(p_1-k\right)^2-M_S^2}\left(p_1^{}+p_2^{}-2k\right)_{}^{\mu}\right]\;,
%     (4)
\end{eqnarray}
and
\begin{eqnarray}
\label{eq:FSb amp}
\overline{u}(p_2^{})\left(-{\rm i} e \Gamma_{\rm FS}^{\mu,({\rm b})}\right)u(p_1^{})
&=&{\rm i} Q_F^{}e \int \frac{{\rm d}^4 k}{\left(2\pi\right)^4}\left[\overline{u}(p_2^{}){\rm i}\left(C_{\rm S}^{*}-C_{\rm P}^{*}\gamma_5^{}\right)\frac{{\rm i}}{\slashed{p}_2^{}-\slashed{k}-M_F^{}}\gamma_{}^{\mu}\right.\nonumber\\
&&\left. \times \frac{{\rm i}}{\slashed{p}_1^{}-\slashed{k}-M_F^{}}{\rm i}\left(C_{\rm S}^{}+C_{\rm P}^{}\gamma_5^{}\right)u(p_1^{})\frac{{\rm i}}{k^2-M_S^2}\right]\;.
%     (5)
\end{eqnarray}
From the overall amplitude, one can extract the FS-type contributions to the muon anomalous magnetic moment
\begin{eqnarray}
\label{eq:FS amu}
a_{\mu}^{{\rm FS}}=\frac{1}{16 \pi^2} f_{\rm FS}\left(\frac{m_{\mu}}{M_S},\frac{M_F}{M_S},C_{\rm S}^{},C_{\rm P}^{}, Q_S^{} \right) \;,
%     (6)
\end{eqnarray}
where the exact relation $Q^{}_F = -1 - Q^{}_S$ has been used and the loop function $f^{}_{\rm FS}$ has been derived without making any approximations. For later convenience, we cast $f^{}_{\rm FS}$ in the following form
\begin{eqnarray}
\label{eq:FS}
f_{\rm FS}^{}\left(x,y,C_{\rm S}^{},C_{\rm P}^{},Q_S \right)=\left(\left| C_{\rm S}^{}\right|_{}^2-\left| C_{\rm P}^{}\right|_{}^2\right)f_{\rm FS}^{-}\left(x,y,Q_S^{}\right)+\left(\left| C_{\rm S}^{}\right|_{}^2+\left| C_{\rm P}^{}\right|_{}^2\right)f_{\rm FS}^{+}\left(x,y,Q_S^{}\right)\;,
%     (7)
\end{eqnarray}
where\footnote{At first glance it seems that the first few terms with negative powers of $x=m_\mu^{}/M_S^{}$ in $f_{\rm FS}^-$ and $f_{\rm FS}^+$ would be divergent as $x$ approaches zero. However, this is not the case since these negative-power terms are exactly cancelled by the relevant terms in the $\Lambda$ function and the final results exhibit only the terms with positive powers of $x$, which would vanish for $m_\mu^{}/M_S^{} \to 0$, as will be shown in Sec.~\ref{subsubsec:hierarchy} below.}
\begin{eqnarray}
\label{eq:FS-}
f_{\rm FS}^{-}\left(x,y,Q_S^{}\right) &\equiv& \frac{2y}{x}+\frac{2y\ln y}{x^3}\left[1-y_{}^2+x_{}^2\left(1+Q_S\right)\right]+\frac{2y\Lambda(x^2,y,1)}{x\lambda(x^2,y^2,1)}\left\{
x_{}^4\left(1+Q_S^{}\right)+Q_S^{}x_{}^2\right.\nonumber\\
&&\left.-\left[2+x_{}^2\left(2+Q_S^{}\right)\right]y_{}^2+y_{}^4
+1\right\}\;,
%     (8)
\end{eqnarray}
and
\begin{eqnarray}
\label{eq:FS+}
f_{\rm FS}^{+}\left(x,y,Q_S^{}\right) &\equiv& -2Q_S^{}-\frac{1}{x^2}\left(2+x^2_{}-2y^2_{}\right)-\frac{2\ln y}{x^4}\left[1-y_{}^2\left(2+x_{}^2-y_{}^2\right)+Q_S^{}x_{}^2\left(1-y_{}^2\right)\right]\nonumber\\
&&+\frac{2 \Lambda(x^2,y,1)}{x^2\lambda(x^2,y^2,1)}\left\{
\left(y_{}^2-1\right)\left[\left(y_{}^2-1\right)_{}^2-x_{}^2\left(1-Q_S^{} \right)-x_{}^2 y_{}^2\left(2+Q_S^{}\right)\right]\right.\nonumber\\
&&\left. +x_{}^4\left[Q_S^{}+y_{}^2\left(1+Q_S^{}\right)\right]
\right\}\;,
%     (9)
\end{eqnarray}
with
\begin{eqnarray}
\Lambda(a,b,c)\equiv \frac{\lambda_{}^{1/2}(a,b_{}^2,c_{}^2)}{a}\ln \left(
\frac{-a+b_{}^2+c_{}^2+\lambda_{}^{1/2}(a,b_{}^2,c_{}^2)}{2 bc}\right)\;
%     (10)
\end{eqnarray}
coming from the branch cut of the Passarino-Veltman $B_0^{}$ function and $\lambda(a,b,c)\equiv a_{}^2 + b_{}^2 + c_{}^2 -2ab -2ac -2bc$ being the K\"all\'en kinematic function. Note that the terms in $f_{\rm FS}^{-}$ $(f_{\rm FS}^{+})$ are all proportional to the odd (even) powers of $y$, reflecting the anti-commutation relation between $\gamma_5^{}$ and $\gamma_{}^{\mu}$. Some helpful comments on the loop function are in order.
\begin{itemize}
\item The general expression of $f_{\rm FS}^{}$ in Eq.~(\ref{eq:FS}) has been written as the sum of two terms, one of which is proportional to $\left(\left|C_{\rm S}^{}\right|_{}^2-\left|C_{\rm P}^{}\right|_{}^2\right)$ and the other to $\left(\left|C_{\rm S}^{}\right|_{}^2+\left|C_{\rm P}^{}\right|_{}^2\right)$. The former term corresponds to the chirality flip via the coupling between $S$ and muon as well as the mass of $F$. This term would vanish if $S$ is not coupled simultaneously to the left- and right-handed muon (i.e., $\left|C_{\rm S}^{}\right|_{}^2-\left|C_{\rm P}^{}\right|_{}^2=0$) or if $F$ is massless (i.e., $M^{}_F = 0$). In contrast, the latter term corresponds to the chirality flip via the SM Yukawa coupling of muon, which is always nonzero since muon is massive. But it may be suppressed, in comparison with the former term, by the mass ratio $m_{\mu}^{}/M_F^{}$.
\item For the electrically neutral $F$ or $S$, only one diagram in Fig.~\ref{fig:FS}(a) or Fig.~\ref{fig:FS}(b) contributes to $a^{}_\mu$. In this case, one can simply substitute $Q_S^{}=-1$ or $Q_S^{}=0$ into Eqs.~(\ref{eq:FS amu})-(\ref{eq:FS+}) to derive the final results.
\end{itemize}

The general formulae of the FS-type contribution to $a^{}_\mu$ in Eqs.~(\ref{eq:FS amu})-(\ref{eq:FS+}) are exact, but it is difficult to get more information from these expressions. Therefore, we make some assumptions on the relevant particle masses and simplify the formulae. As will be shown below, some interesting results can be obtained.

\subsubsection{$m_{\mu}^{}\ll M_F^{}, M_S^{}$}
\label{subsubsec:hierarchy}
The first scenario is to assume that both $F$ and $S$ are particles much heavier than muon, i.e., $m_{\mu}^{}\ll M_F^{}, M_S^{}$. Note that it is not necessary to specify the relative size of $M^{}_F$ and $M^{}_S$, which can actually be comparable. In terms of the small mass ratio $m_{\mu}^{}/M_S^{}$, we can expand the functions $f^\mp_{\rm FS}$ in Eqs.~(\ref{eq:FS-}) and (\ref{eq:FS+}) as follows
\begin{eqnarray}
\label{eq:FS series}
f_{\rm FS}^{} &=& +\left(\left|C_{\rm S}^{}\right|_{}^2-\left|C_{\rm P}^{}\right|_{}^2\right) \sum_{n=1}^{\infty}g_{\rm \rm FS}^{(2n-1)} \left(\frac{M^{}_F}{M^{}_S},Q_S^{} \right) \left(\frac{m^{}_{\mu}}{M^{}_S}\right)_{}^{2n-1} \nonumber\\
&& + \left(\left|C_{\rm S}^{}\right|_{}^2+\left|C_{\rm P}^{}\right|_{}^2\right)\sum_{n=1}^{\infty}g_{\rm FS}^{(2n)} \left(\frac{M^{}_F}{M^{}_S},Q^{}_S\right) \left(\frac{m^{}_{\mu}}{M^{}_S}\right)_{}^{2n}\;,
%     (11)
\end{eqnarray}
where the terms of odd (even) powers come from $f_{\rm FS}^{-}$ ($f_{\rm FS}^{+}$). In light of $m^{}_\mu/M^{}_S \ll 1$, it is an excellent approximation to keep only the first two leading terms in Eq.~(\ref{eq:FS series}), i.e., the first- and second-order loop functions 
\begin{eqnarray}
\label{eq:gFS1}
g_{\rm FS}^{(1)}\left(y,Q_S^{}\right) = \frac{y\left[
y_{}^4\left(1+2Q_S^{}\right)-4y_{}^2\left(1+Q_S^{}+Q_S^{}\ln y\right) + 4\left(1+Q_S^{}\right)\ln y + 3 + 2Q_S^{}
\right]}{\left(y^2-1\right)^3}\;,
%     (12)
\end{eqnarray}
and
\begin{eqnarray}
\label{eq:gFS2}
g_{\rm FS}^{(2)}\left(y,Q_S^{}\right) &=& \frac{1}{3\left(y^2-1\right)^4}\left\{y_{}^6\left(1+3Q_S^{}\right) - 3y_{}^4\left(2+Q_S^{}+4Q_S^{}\ln y\right)\right. \nonumber\\
&&\left. +3y_{}^2\left[1-Q_S^{}+4\left(1+Q_S^{}\right)\ln y\right] +2+3Q_S^{}\right\}\;.
%     (13)
\end{eqnarray}
%%%%%%%%%%%%%%%%%%%%%%%%%%%%%%%%%%%% Fig 2 
%%%%%%%%%%%%%%%%%%%%%%%%%%%%%%%
\begin{figure}[t!]
	\centering		
	\subfigure[The first-order loop function $g_{\rm FS}^{(1)}\left(y,Q_S^{}\right)$ for $Q_S^{}=-1$ (red dashed line), $Q_S^{}=-2/3$ (gray solid line) and $Q_S^{}=-1/2$ (blue dash-dotted line). Note that for $Q_S^{}\geq -1/2$, $g_{\rm FS}^{(1)}\left(y,Q_S^{}\right)$ is non-negative; for $Q_S^{}\leq -1$, $g_{\rm FS}^{(1)}\left(y,Q_S^{}\right)$ is non-positive; for $-1<Q_S^{}<-1/2$, the sign of $g_{\rm FS}^{(1)}\left(y,Q_S^{}\right)$ is indefinite.]{\includegraphics[width=0.48\linewidth]{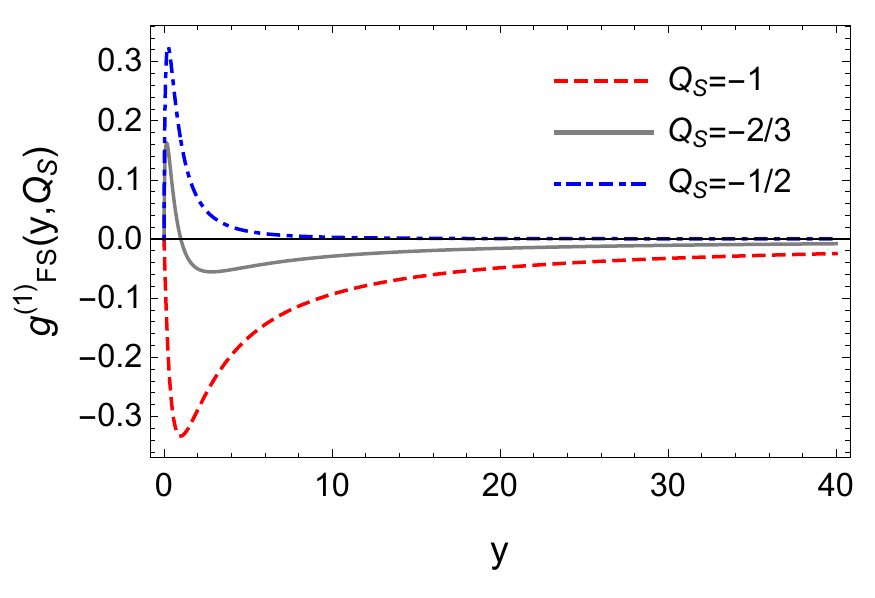}}\quad
	\subfigure[The second-order loop function $g_{\rm FS}^{(2)}\left(y,Q_S^{}\right)$ for $Q_S^{}=-2/3$ (red dashed line), $Q_S^{}=-1/2$ (gray solid line) and $Q_S^{}=-1/3$ (blue dash-dotted line). Note that for $Q_S^{}\geq -1/3$, $g_{\rm FS}^{(2)}\left(y,Q_S^{}\right)$ is non-negative; for $Q_S^{}\leq-2/3$, $g_{\rm FS}^{(2)}\left(y,Q_S^{}\right)$ is non-positive; for $-2/3<Q_S^{}<-1/3$, the sign of $g_{\rm FS}^{(2)}\left(y,Q_S^{}\right)$ is indefinite.]{\includegraphics[width=0.48\linewidth]{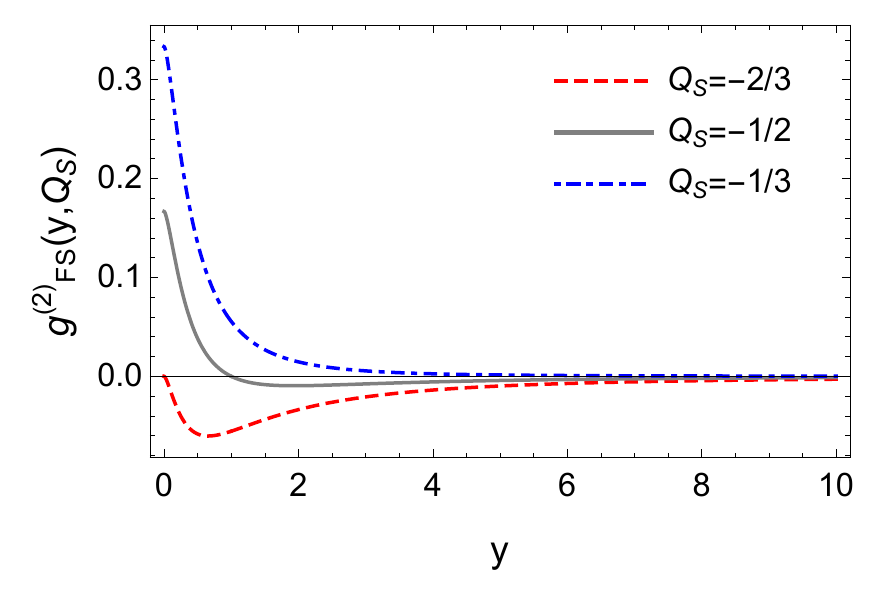}}
	\vspace{0.5cm}
	\caption{Numerical illustrations for the first- and second-order loop functions relevant for the FS-type contributions to $a^{}_\mu$ for different values of the electric charge $Q_S^{}$.}
	\label{fig:gFS}
\end{figure}
%%%%%%%%%%%%%%%%%%%%%%%%%%%%%%%%%%%%%%%%%%%%%%%%%%%%%%%%%%%%%%%%%%%%%%%%%%

Based on the baisc properties of the loop functions $g_{\rm FS}^{(n)}(y,Q_S^{})$ with $n$ being positive integers, we can make a few interesting observations. 
\begin{itemize}
\item First of all, one may wonder whether the loop functions $g_{\rm FS}^{(1)}(y,Q_S^{})$ and $g_{\rm FS}^{(2)}(y,Q_S^{})$ in Eqs.~(\ref{eq:gFS1}) and (\ref{eq:gFS2}) become divergent at the point $y = 1$, i.e., $M^{}_F = M^{}_S$. After a direct calculation, we verify that all the functions $g^{(n)}_{\rm FS}(y, Q^{}_S)$ are actually finite and continuous at $y=1$, implying the absence of a singularity at $M_F^{}=M_S^{}$.

\item Second, all the loop functions $g^{(n)}_{\rm FS}(y, Q^{}_S)$ tend to be zero as $y$ is approaching infinity (i.e., in the limit $M_F^{}\gg M_S^{}$). On the other hand, if $y$ is approaching zero (i.e., in the limit $M_F^{}\ll M_S^{}$), then we have $g_{\rm FS}^{(n)}(y, Q^{}_S)$ with $n$ being odd tends to be zero, as a consequence of the fact that these contributions come from the chirality flip via the mass of $F$. Moreover, the loop functions $g_{\rm FS}^{(n)}(y, Q^{}_S)$ with $n$ being even are finite in the limit of $M_F^{}\ll M_S^{}$ or $y = 0$, and the finite value depends on $n$ and $Q_S^{}$.

\item Third, we emphasize whether the loop functions $g_{\rm FS}^{(n)}(y, Q^{}_S)$ at any values of $y$ are positive or negative depends sensitively on the electric charge $Q_S^{}$. In general, there exist both an upper bound $Q^{\rm U}_S$ and a lower bound $Q^{\rm L}_S$ for the electric charge $Q^{}_S$ such that $g_{\rm FS}^{(n)}(y, Q^{}_S) \geq 0$ for $Q^{}_S \geq Q^{\rm U}_S$ and $g_{\rm FS}^{(n)}(y, Q^{}_S) \leq 0$ for $Q^{}_S \leq Q^{\rm L}_S$ hold. However, for $Q_{S}^{\rm L} < Q_S^{} < Q_{S}^{\rm U}$, the sign of the loop functions will be indefinite. The exact values of the corresponding $Q_S^{\rm U}$ and $Q_{S}^{\rm L}$ vary with $n$. 
\end{itemize}
%%%%%%%%%%%%%%%%%%%%%%%%%%%%%%%%%%% Table 1 %%%%%%%%%%%%%%%%%%%%%%%%%%%%%%
\begin{table}[t!]
\centering
\renewcommand\arraystretch{2.5}
\begin{tabular}{c|c|c|c|c|c}
\hline \hline
{\bf Loop functions} & $y\to 0$ & $y\to \infty$ & $y \to 1$ & $Q_{S}^{\rm L}$ & $Q_S^{\rm U}$ \\
\hline\hline
 $g_{\rm FS}^{(1)}\left(y,Q_S^{}\right)$ & 0 & 0 &  $\displaystyle \frac{2}{3}+Q_S^{}$ & $-1$ & $\displaystyle -\frac{1}{2}$
 \\
\hline
$g_{\rm FS}^{(2)}\left(y,Q_S^{}\right)$ & $\displaystyle \frac{2}{3}+Q_S^{}$ & 0 & $\displaystyle \frac{1}{6}\left(1+2Q_S^{}\right)$ & $\displaystyle -\frac{2}{3}$ & $\displaystyle -\frac{1}{3}$
\\
\hline\hline
\end{tabular}
\vspace{0.5cm}
\caption{Summary of the asymptotic values of $g^{(1)}_{\rm FS}(y, Q^{}_S)$ and $g^{(2)}_{\rm FS}(y, Q^{}_S)$ in the limits of $y \to 0$, $y \to \infty$ and $y \to 1$, as well as the corresponding critical values $Q^{\rm U}_S$ and $Q^{\rm L}_S$.}
\label{table:FS}
\end{table}
\renewcommand\arraystretch{1}
%%%%%%%%%%%%%%%%%%%%%%%%%%%%%%%%%%%%%%%%%%%%%%%%%%%%%%%%%%%%%%%%%%%%%%%%%%

For illustration we focus on the first- and second-order loop functions $g^{(1)}_{\rm FS}(y, Q^{}_S)$ and $g^{(2)}_{\rm FS}(y, Q^{}_S)$ in Eqs.~(\ref{eq:gFS1}) and (\ref{eq:gFS2}), and show them in Fig.~\ref{fig:gFS} for different values of $Q_S^{}$. The critical values $Q^{\rm U}_S = -1/2$ and $Q^{\rm L}_S = -1$ have been found for $g^{(1)}_{\rm FS}(y, Q^{}_S)$, while $Q^{\rm U}_S = -1/3$ and $Q^{\rm L}_S = -2/3$ for $g^{(2)}_{\rm FS}(y, Q^{}_S)$. In addition, the asymptotic values of $g^{(1)}_{\rm FS}(y, Q^{}_S)$ and $g^{(2)}_{\rm FS}(y, Q^{}_S)$ in the limits of $y \to 0$, $y \to \infty$ and $y \to 1$, together with the critical values $Q^{\rm U}_S$ and $Q^{\rm L}_S$, have been summarized in Table~\ref{table:FS}. These observations are useful in judging whether the FS-type contribution to $a^{}_\mu$ is positive or negative. For instance, for the pure left- or right- handed Yukawa-type interactions, we have $|C^{}_{\rm S}|^2 - |C^{}_{\rm P}|^2 = 0$ and thus the dominant contribution to $a^{}_\mu$ is given by $g^{(2)}_{\rm FS}(y, Q^{}_S)$ in Eq.~(\ref{eq:FS series}). In this case, one can conclude that $\Delta a^{\rm FS}_\mu > 0$ for $Q^{}_S \geq -1/3$ and $\Delta a^{\rm FS}_\mu < 0$ for $Q^{}_S \leq -2/3$.

\subsubsection{$m_{\mu}^{}, M_F^{} \ll M_S^{}$}

Then we proceed with another scenario where $S$ is much heavier than $F$ and muon, i.e., $m_{\mu}^{}, M_F^{} \ll M_S^{}$. To the order of ${\cal O}\left(m^2_{\mu}/M_S^2\right)$, one obtains
\begin{eqnarray}
\label{eq:FS appro2}
f^{}_{\rm FS} &=& \frac{1}{3}\left(\frac{m_\mu^{}}{M_S^{}}\right)_{}^2\left\{\left(\left| C_{\rm S}^{}\right|_{}^2+\left| C_{\rm P}^{}\right|_{}^2\right)\left(2+3Q_S^{}\right)-3 \left(\left| C_{\rm S}^{}\right|_{}^2-\left| C_{\rm P}^{}\right|_{}^2\right) \frac{M^{}_F}{m^{}_\mu} \right.\nonumber\\
&&\left. \times \left[3+2Q_S^{} +4\left(1+Q_S^{}\right)\ln\left(\frac{M_F^{}}{M_S^{}}\right) \right]
\right\}\;,
%     (14)
\end{eqnarray}
where the mass ratio $M^{}_F/m^{}_\mu$ can take any positive values. In the limit of $M_F^{}/m_{\mu}^{}\to 0$, which is valid for $F$ being neutrinos in the SM, the loop function in Eq.~(\ref{eq:FS appro2}) is reduced to
\begin{eqnarray}
\label{eq:FS appro2p}
f_{\rm FS}^{}=\frac{1}{3}\left(\frac{m_{\mu}^{}}{M_S^{}}\right)^2_{}\left(\left| C_{\rm S}^{}\right|_{}^2+\left| C_{\rm P}^{}\right|_{}^2\right)\left(2+3Q_S^{}\right)\;,
%     (15)
\end{eqnarray}
where the term proportional to $\left(\left| C_{\rm S}^{}\right|_{}^2-\left| C_{\rm P}^{}\right|_{}^2\right)$ vanishes because of the fact that the chirality flip via the mass of $F$ becomes impossible in the limit $M^{}_F/m^{}_\mu \to 0$. As indicated by Eq.~(\ref{eq:FS appro2p}), in this case, the contribution to $a^{}_\mu$ is positive only for $Q^{}_S > -2/3$. 

If the mass hierarchy $m_\mu^{}\ll M_F^{}$ is further assumed, then Eq.~(\ref{eq:FS appro2}) can be reduced to
\begin{eqnarray}
\label{eq:FS appro2pp}
f_{\rm FS}^{}=-\frac{m_\mu M_F}{M_S^2}\left(\left| C_{\rm S}^{}\right|_{}^2-\left| C_{\rm P}^{}\right|_{}^2\right)\left[3+2Q_S^{} +4\left(1+Q_S\right){\rm ln}\left(\frac{M_F}{M_S}\right)\right]\; .
\end{eqnarray}
It is worthwhile to mention that the result in Eq.~(\ref{eq:FS appro2pp}) can also be obtained by taking $y\equiv M_F^{}/M_S^{}\ll 1$ in the function $g_{\rm FS}^{(1)}(y,Q_S^{})$ in Eq.~(\ref{eq:gFS1}).

\subsubsection{$m_{\mu}^{}, M_S^{}\ll M_F^{}$}
\label{subsec:mmu,MS<<MF}
Finally we consider the scenario where both muon and $S$ are much lighter than $F$, i.e., $m_{\mu}^{}, M_S^{}\ll M_F^{}$. To the order of ${\cal O}\left(m_{\mu}^2/M_F^2\right)$, we have
\begin{eqnarray}
\label{eq:FS appro3}
f_{\rm FS}^{}=\left(\left| C_{\rm S}^{}\right|_{}^2-\left| C_{\rm P}^{}\right|_{}^2\right)\left(1+2Q_S^{}\right)\left(\frac{m_\mu}{M_F}\right)+\frac{1}{3}\left(\left| C_{\rm S}^{}\right|_{}^2+\left| C_{\rm P}^{}\right|_{}^2\right)
\left(1+3Q_S^{}\right)\left(\frac{m_\mu}{M_F}\right)_{}^2\;,
%     (16)
\end{eqnarray}
which can also be obtained by taking the limit of $y=M_F^{}/M_S^{}\rightarrow \infty$ in Eqs.~(\ref{eq:gFS1}) and (\ref{eq:gFS2}). Note that to the leading and next-to-leading orders of $m^{}_\mu/M^{}_F$, the contributions to $a_\mu^{}$ have nothing to do with the mass of $S$, as long as $F$ is much heavier than muon and $S$. In particular, we shall have a finite result even for a massless $S$. This is very different from the FV-case where there is a vector boson instead of a scalar boson in the loop, as we will see in the next subsection, the loop function would be divergent as the mass of the vector boson were approaching zero.

\subsection{The FV-type}
\label{sec:F+V}
%%%%%%%%%%%%%%%%%%%%%%%%%%%%%%% Fig. 3 %%%%%%%%%%%%%%%%%%%%%%%%%%%%%%%%
\begin{figure}[t!]
	\centering		\includegraphics[width=17.5cm]{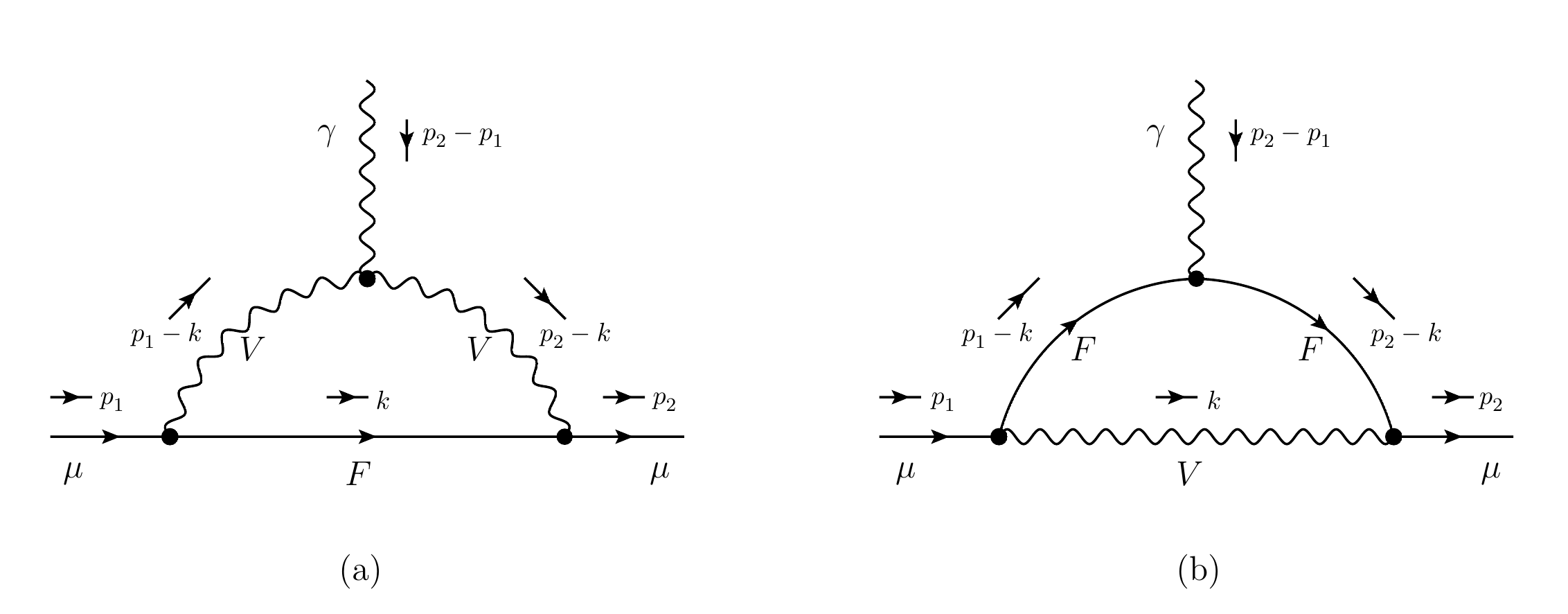}
	\vspace{-0.8cm}
	\caption{The Feynman diagrams for the FV-type contributions to the muon anomalous magnetic moment $a^{}_\mu$, where the particle four-momenta have been explicitly specified and represented by arrows along the internal and external lines.}
	\label{fig:FV}
\end{figure}
%%%%%%%%%%%%%%%%%%%%%%%%%%%%%%%%%%%%%%%%%%%%%%%%%%%%%%%%%%%%%%%%%%%%%%%
Then we turn to the case where there is a fermion $F$ and a vector boson $V$ in the loops that contribute to $a_\mu^{}$. The relevant Feynman diagrams are shown in Fig.~\ref{fig:FV}. These FV-type diagrams actually appear in some realistic models, such as the $W$- and $Z$-boson contributions in the SM and the type-I seesaw model, the $Z_{}^{\prime}$ model, and the vector leptoquark model.

After the spontaneous breaking of the ${\rm SU}(2)_{\rm L}^{}\otimes {\rm U}(1)_{\rm Y}^{}$ gauge symmetry, the interactions among all the relevant particles can be written as
\begin{eqnarray}
\label{eq:FV langrangian}
{\cal L}_{\rm int}^{{\rm FV}}&=&\left[\overline{F}\left(C_{\rm V}^{}\gamma_{}^{\mu}+C_{\rm A}^{}\gamma_{}^{\mu}\gamma_5^{} \right)\mu V_{\mu}^{\dagger}+{\rm h.c.}\right]+Q_F^{}e\overline{F}\slashed{A}F +{\rm i} e Q_V^{}\left[\left(\partial_{\mu}^{}V_{\nu}^{\dagger}-\partial_{\nu}^{}V_{\mu}^{\dagger}\right)A_{}^{\mu}V_{}^{\nu}\right.\nonumber\\
&& \left. -\left(\partial_{\mu}^{}V_{\nu}^{}-\partial_{\nu}^{}V_{\mu}^{}\right)A_{}^{\mu}V_{}^{\dagger \nu}+\left(V_{\mu}^{}V_{\nu}^{\dagger}-V_{\nu}^{}V_{\mu}^{\dagger} \right)\partial_{}^{\mu}A_{}^{\nu}
\right]\;,
%     (17)
\end{eqnarray}
where $C_{\rm V}^{}$ and $C_{\rm A}^{}$ stand respectively for the vector-type and axial-vector-type coupling constants. The interaction between $V$ (or $F$) and photon is determined by the minimal coupling as in the ordinary QED. The last two terms $\left(V_{\mu}^{}V_{\nu}^{\dagger}-V_{\nu}^{}V_{\mu}^{\dagger} \right)\partial_{}^{\mu}A_{}^{\nu}$ in Eq.~(\ref{eq:FV langrangian}) have been added to ensure the renormalizability of the theory~\cite{Biggio2016}. In addition, the conservation of electric charges requires the relation $Q_F^{}+Q_V^{}=-1$ to hold.

The amplitudes corresponding to the Feynman diagrams in Fig.~\ref{fig:FV}(a) and Fig.~\ref{fig:FV}(b) can be explicitly written down as below\footnote{Throughout this paper, whenever the vector boson $V$ is involved, we perform all the calculations in the unitary gauge so that $V$ can be either a gauge boson or a vector-like matter particle.}
\begin{eqnarray}
\label{eq:FVa amp}
\overline{u}(p_2^{})\left(-{\rm i} e \Gamma_{\rm FV}^{\mu,({\rm a})}\right)u(p_1^{})
&=&{\rm i} Q_V^{}e \int \frac{{\rm d}^4 k}{\left(2\pi\right)^4}\left\{\overline{u}(p_2^{}){\rm i}\left(C_{\rm V}^{*}\gamma_{}^{\sigma}+C_{\rm A}^{*}\gamma_{}^{\sigma}\gamma_5^{}\right)\frac{{\rm i}}{\slashed{k}-M_F^{}}{\rm i}\left(C_{\rm V}^{}\gamma_{}^{\rho}\gamma_5^{}+C_{\rm A}^{}\gamma_{}^{\rho}\gamma_5^{}\right)\right.\nonumber\\
&&\left. \times u(p_1^{})
\frac{{-\rm i}}{\left(p_1-k\right)^2-M_V^2}\frac{{-\rm i}}{\left(p_2-k\right)^2-M_V^2}\left[\eta_{\rho \alpha}-\frac{(p_1-k)_\rho (p_1-k)_\alpha}{M_V^2}\right]\right.\nonumber\\
&& \left. \times \left[\eta_{\sigma \beta}-\frac{(p_2-k)_\sigma (p_2-k)_\beta}{M_V^2}\right]\left[\eta_{}^{\beta\alpha}\left(2k-p_1^{}-p_2^{}\right)_{}^{\mu}\right.\right.\nonumber\\
&&\left.\left. +\eta_{}^{\alpha\mu}\left(2p_1^{}-p_2^{}-k\right)_{}^{\beta} +\eta_{}^{\mu\beta}\left(2p_2^{}-k-p_1^{}\right)_{}^{\alpha}
\right]
\right\}\;,
%     (18)
\end{eqnarray}
and
\begin{eqnarray}
\label{eq:FVb amp}
\overline{u}(p_2^{})\left(-{\rm i} e \Gamma_{\rm FV}^{\mu,({\rm b})}\right)u(p_1^{})
&=&{\rm i} Q_F^{}e \int \frac{{\rm d}^4 k}{\left(2\pi\right)^4}\left[\overline{u}(p_2^{}){\rm i}\left(C_{\rm V}^{*}\gamma_{}^{\sigma}+C_{\rm A}^{*}\gamma_{}^{\sigma}\gamma_5^{}\right)\frac{{\rm i}}{\slashed{p}_2^{}-\slashed{k}-M_F^{}}\gamma_{}^{\mu}\right.\nonumber\\
&&\left. \times \frac{{\rm i}}{\slashed{p}_1^{}-\slashed{k}-M_F^{}}{\rm i}\left(C_{\rm V}^{}\gamma_{}^{\rho}+C_{\rm A}^{}\gamma_{}^{\rho}\gamma_5^{}\right)u(p_1^{})\frac{-{\rm i}}{k^2-M_V^2}\left(\eta_{\rho \sigma}-\frac{k_{\rho}^{}k_{\sigma}^{}}{M_V^2}\right)\right]\;,\nonumber\\
%     (19)
\end{eqnarray}
where $\eta^{}_{\mu \nu}={\rm diag}(+,-,-,-)$ is the metric tensor of the Minkowski space-time. From Eqs.~(\ref{eq:FVa amp}) and (\ref{eq:FVb amp}) one can extract the FV-type contributions to $a^{}_\mu$, i.e.,
\begin{eqnarray}
\label{eq:FV amu}
a_{\mu}^{\rm FV}=\frac{1}{16 \pi^2}f_{\rm FV}\left(\frac{m_{\mu}}{M_V},\frac{M_F}{M_V},C_{\rm V}^{},C_{\rm A}^{}, Q_V^{} \right)\;,
%     (20)
\end{eqnarray}
where the exact relation $Q_F^{} =-1-Q_V^{}$ has been used and the loop function $f_{\rm FV}^{}$ has been derived without making any approximations. Similar to $f^{}_{\rm FS}$ in the previous subsection, $f_{\rm FV}^{}$ can also be divided into two parts
\begin{eqnarray}
\label{eq:FV}
f_{\rm FV}^{}\left(x,y,C_{\rm V}^{},C_{\rm A}^{},Q_S \right)=\left(\left| C_{\rm V}^{}\right|_{}^2-\left| C_{\rm A}^{}\right|_{}^2\right)f_{\rm FV}^{-}\left(x,y,Q_V^{}\right)+\left(\left| C_{\rm V}^{}\right|_{}^2+\left| C_{\rm A}^{}\right|_{}^2\right)f_{\rm FV}^{+}\left(x,y,Q_V^{}\right)\;,
%     (21)
\end{eqnarray}
where
\begin{eqnarray}
\label{eq:FV-}
f_{\rm FV}^{-}\left(x,y,Q_V^{}\right)&=&
4 Q_V^{}x y-\frac{2y}{x}\left(2+y_{}^2-2x_{}^2\right)-\frac{2y\ln y}{x^3}\left\{2+x_{}^2\left(2Q_V-1\right)-x_{}^4\left(1+Q_V^{}\right)\right.\nonumber\\
&&\left. +y_{}^2\left[\left(2+Q_V^{}\right)x_{}^2-1\right]-y_{}^4 \right\}
-\frac{2y \Lambda(x^2_{},y,1)}{x\lambda(x^2_{},y^2_{},1)}\left\{2-x_{}^2\left[3+x_{}^4 +Q_V^{}\left(-2+3x_{}^2\right.\right.\right.\nonumber\\
&&\left.\left.\left. +x_{}^4\right) \right]+y_{}^2\left[-3-Q_V^{}x_{}^2+x_{}^4\left(3+2Q_V^{}\right)\right]-x_{}^2y_{}^4\left(3+Q_V^{}\right)+y_{}^6\right\}\;,
%     (22)
\end{eqnarray}
and
\begin{eqnarray}
\label{eq:FV+}
f_{\rm FV}^{+}\left(x,y,Q_V^{}\right)&=&
-2Q_V^{}\left(2+x_{}^2+y_{}^2\right)-\frac{1}{x^2}\left[\left(x_{}^2-2\right)_{}^2+3x_{}^2y_{}^2-2y_{}^2\left(1+y_{}^2\right) \right]-\frac{2 \ln y}{x^4}\nonumber\\
&&\times \left\{2+x_{}^2\left(2 Q_V^{}-3-3Q_V^{}x_{}^2\right)+y_{}^2\left[ -3+x_{}^2\left(2-Q_V^{}\right)+x_{}^4\left(1+Q_V^{}\right)\right]\right.\nonumber\\
&&\left. -x_{}^2y_{}^4\left(2+Q_V^{}\right)+y_{}^6\right\}-\frac{2 \Lambda(x^2_{},y,1)}{x^2\lambda(x^2_{},y^2_{},1)}\left\{\left(1+Q_V^{}x_{}^2\right)\left(2-5x_{}^2+3x_{}^4\right)\right.\nonumber\\
&&\left. + y_{}^2\left[-5+x_{}^2\left(4+x_{}^2+x_{}^4+Q_V^{}\left(-3+x_{}^2+x_{}^4 \right)\right)\right]\right.\nonumber\\
&& \left. +y_{}^4\left[3-2x_{}^2-x_{}^4\left(3+2Q_V^{}\right)\right] +y_{}^6\left[1+x_{}^2\left(3+Q_V^{}\right)\right]-y_{}^8
\right\}\;.
%     (23)
\end{eqnarray}

Although the expressions of the loop functions in Eqs.~(\ref{eq:FV-}) and (\ref{eq:FV+}) seem to be lengthy, the integration has been explicitly carried out and the results are exact in the sense that no approximations have been made. As in the case of FS-type contributions, the anti-commutation relation between $\gamma_5^{}$ and $\gamma_{}^{\mu}$ enforces the terms in $f_{\rm FV}^{-}$ (or $f_{\rm FV}^{+}$) to be proportional to the odd (or even) powers of $y$. Some comments on the loop functions are in order.
\begin{itemize}
\item The loop function $f_{\rm FV}^{-}$ corresponds to the chirality flips via the coupling between $V$ and muon or via the mass of $F$, which would vanish if $V$ were not coupling simultaneously to the left- and right- handed muon (i.e., $\left|C_{\rm V}^{}\right|_{}^2-\left|C_{\rm A}^{}\right|_{}^2=0$) or if $F$ were  massless (i.e., $M_F^{}=0$). In contrast, the other loop function $f_{\rm FV}^{+}$ corresponds to the chirality flips via the SM Yukawa coupling of muon, which is always nonzero because of non-vanishing muon mass. But it may be suppressed, when compared to $f_{\rm FV}^{-}$, by the mass ratio $m_{\mu}^{}/M_F^{}$.

\item If $F$ or $V$ is electrically neutral, then only the diagram in Fig.~\ref{fig:FV}(a) or Fig.~\ref{fig:FV}(b) contributes to $a_\mu^{}$. In this case, one can simply set $Q_V^{}=-1$ or $Q_V^{}=0$ in Eqs.~(\ref{eq:FV amu})-(\ref{eq:FV+}) to derive the final results.
\end{itemize}

In the remaining part of this subsection we shall impose some hierarchical conditions on the masses of the relevant particles to simplify the general expressions of Eqs.~(\ref{eq:FV-}) and (\ref{eq:FV+}).

\subsubsection{$m_{\mu}^{}\ll M_F^{}, M_V^{}$}
\label{subsec:mmu<< MF,MV}
Suppose that both $F$ and $V$ are much heavier than muon, i.e., $m_{\mu}^{}\ll M_F^{}, M_V^{}$, whereas the relative sizes of $M_F^{}$ and $M_V^{}$ are not specified. Expanding the general functions $f_{\rm FV}^{\mp}$ in Eqs.~(\ref{eq:FV-}) and (\ref{eq:FV+}) as the series of $m_{\mu}^{}/M_V^{}$, one obtains
\begin{eqnarray}
\label{eq:FV series}
f_{\rm FV}^{}&=&+\left(\left|C_{\rm V}^{}\right|_{}^2-\left|C_{\rm A}^{}\right|_{}^2\right)\sum_{n=1}^{\infty}g_{\rm FV}^{(2n-1)}\left(\frac{M_F}{M_V},Q_V^{} \right)\left(\frac{m_{\mu}}{M_V}\right)_{}^{2n-1}\nonumber\\
&&+\left(\left|C_{\rm V}^{}\right|_{}^2+\left|C_{\rm A}^{}\right|_{}^2\right)\sum_{n=1}^{\infty}g_{\rm FV}^{(2n)}\left(\frac{M_F}{M_V},Q_V^{} \right)\left(\frac{m_{\mu}}{M_V}\right)_{}^{2n}\;,
%     (24)
\end{eqnarray}
where the odd- and even-power terms come from $f_{\rm FV}^{-}$ and $f_{\rm FV}^{+}$, respectively. In the assumption of $m_\mu^{}\ll M_V^{}$, it is a good approximation to keep the first two leading terms in Eq.~(\ref{eq:FV series}), i.e., the first- and second-order loop functions
\begin{eqnarray}
\label{eq:gFV1}
g_{\rm FV}^{(1)}\left(y,Q_V^{}\right)&=&
\frac{y}{(y^2-1)^3}\left\{
y_{}^6\left(1+2Q_V^{}\right)-12y_{}^4 Q_V^{}\left(1-\ln y\right)+3y_{}^2\left[1+6Q_V^{}-4\left(1+Q_V^{}\right)\ln y\right]\right.\nonumber\\
&& \left. -4\left(1+2Q_V^{}\right)
\right\}\;,
%     (25)
\end{eqnarray}
and
\begin{eqnarray}
\label{eq:gFV2}
g_{\rm FV}^{(2)}\left(y,Q_V^{}\right)&=&
\frac{-1}{3\left(y^2-1\right)^4}\left\{y_{}^8\left(5+9 Q_V^{}\right)-y_{}^6\left[7\left(2+9Q_V^{}\right)-36 Q_V^{}\ln y\right]\right.\nonumber\\
&&\left. +y_{}^4\left[39+117 Q_V^{}-36\left(1+Q_V^{}\right)\ln y\right]-y_{}^2\left(38+81 Q_V^{}\right)+2\left(4+9Q_V^{}\right)
\right\}
\;. \quad
%     (26)
\end{eqnarray}

\begin{figure}[t!]
	\centering		
	\subfigure[The first-order loop function $g_{\rm FV}^{(1)}\left(y,Q_V^{}\right)$ for $Q_V^{}=-1/2$ (red dashed line), $Q_V^{}=-2/5$ (gray solid line) and $Q_V^{}=-1/3$ (blue dash-dotted line). Note that for $Q_V^{}\geq -0.387$, $g_{\rm FV}^{(1)}\left(y,Q_V^{}\right)$ is non-negative; for $Q_V^{}\leq -1/2$, $g_{\rm FV}^{(1)}\left(y,Q_V^{}\right)$ is non-positive; for $-1/2<Q_S^{}<-0.387$, the sign of $g_{\rm FV}^{(1)}\left(y,Q_S^{}\right)$ is indefinite.]{\includegraphics[width=0.48\linewidth]{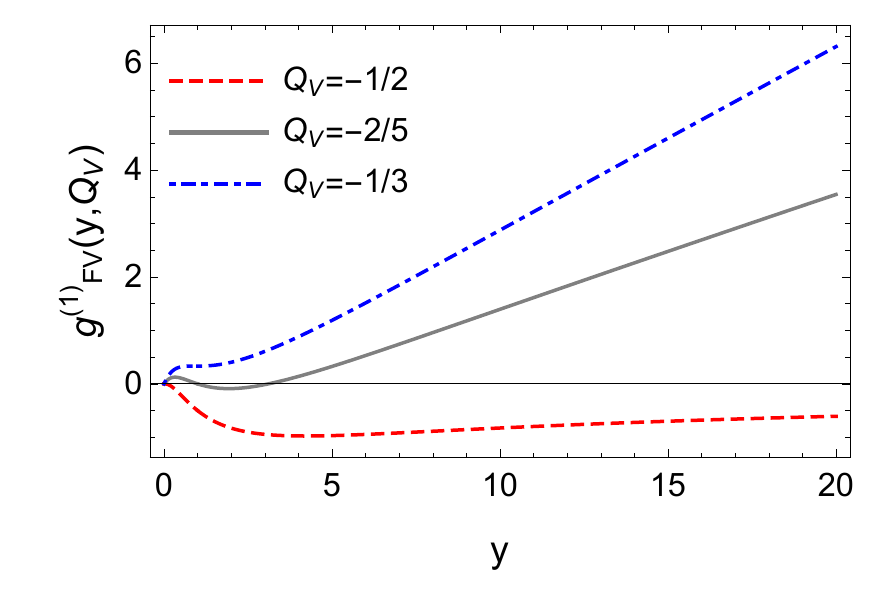}}\quad
	\subfigure[The second-order loop function $g_{\rm FV}^{(2)}\left(y,Q_V^{}\right)$ for $Q_V^{}=-5/9$ (red dashed line), $Q_V^{}=-1/2$ (gray solid line) and $Q_V^{}=-1/3$ (blue dash-dotted line). Note that for $Q_V^{}\geq-0.433$, $g_{\rm FV}^{(2)}\left(y,Q_V^{}\right)$ is non-positive; for $Q_V^{}\leq-5/9$, $g_{\rm FV}^{(2)}\left(y,Q_V^{}\right)$ is non-negative; for $-5/9<Q_V^{}<-0.433$, the sign of $g_{\rm FV}^{(2)}\left(y,Q_V^{}\right)$ is indefinite.]{\includegraphics[width=0.48\linewidth]{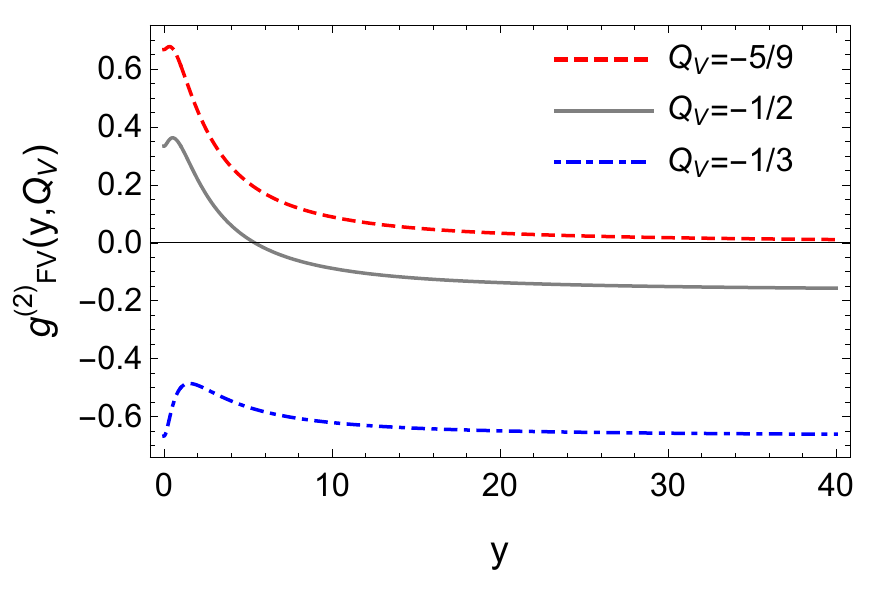}}
	\vspace{0.5cm}
	\caption{Numerical illustrations for the first- and second-order loop functions relevant for the FV-type contributions to $a_\mu^{}$ for different values of the electric charge $Q_V^{}$.}
	\label{fig:gFV}
\end{figure}

Some interesting observations on the basic properties of the loop functions $g_{\rm FV}^{(n)}(y,Q_V^{})$ (with $n$ being positive integers) can be made. 
\begin{itemize}
\item All the loop functions $g_{\rm FV}^{(n)}(y,Q_V^{})$ are finite and continuous at $y=1$, implying the absence of a singularity at $M_V^{}=M_F^{}$. As $y$ is approaching zero (i.e., in the limit $M_F^{}\ll M_V^{}$), $g_{\rm FV}^{(n)}(y,Q_V^{})$ with $n$ being odd tends to be zero, reflecting the fact that these contributions arise from the chirality flip via the mass of $F$, while $g_{\rm FV}^{(n)}(y,Q_V^{})$ with $n$ being even has a finite limit that depends on both $n$ and $Q_V^{}$. When $y$ is approaching infinity (i.e., in the limit $M_F^{}\gg M_V^{}$), $g_{\rm FV}^{(n)}(y,Q_V^{})$ with $n>2$ becomes zero. However, in the same limit, we have $g_{\rm FV}^{(2)}(y,Q_V^{}) \to -5/3-3Q_V^{}$, while $g_{\rm FV}^{(1)}(y,Q_V^{}) \to \infty$ for $Q_V^{} \neq -1/2$ and $g_{\rm FV}^{(1)}(y,Q_V^{}) \to 0$ for $Q_V^{} = -1/2$.

\item There exist both an upper bound $Q_V^{\rm U}$ and a lower bound $Q_V^{\rm L}$ for the electric charge $Q_V^{}$ such that
$g_{\rm FV}^{(n)}(y,Q_V^{})\geq 0$ ($\leq 0$) for $Q_V^{}\geq Q_V^{\rm U}$ with $n$ being odd (even) and $g_{\rm FV}^{(n)}(y,Q_V^{})\leq 0$ ($\geq 0$) for $Q_V^{}\leq Q_V^{\rm L}$ with $n$ being odd (even). However, for $Q_V^{\rm L}<Q_V^{}<Q_V^{\rm U}$, the sign of the loop functions will be indefinite. The exact values of the corresponding $Q_V^{\rm U}$ and $Q_V^{\rm L}$ vary with $n$. For a better illustration we have shown the first- and second-order loop functions $g_{\rm FV}^{(1)}(y,Q_V^{})$ and $g_{\rm FV}^{(2)}(y,Q_V^{})$ in Fig.~\ref{fig:gFV} for different values of $Q_V^{}$. The asymptotic behaviors  together with the critical values $Q_V^{\rm U}$ and $Q_V^{\rm L}$ are also summarized in Table~\ref{table:FV}. This observation is useful in judging the sign of the FV-type contribution to $a_\mu^{}$. For instance, for the pure left- or right-handed interaction, we have $\left|C_{\rm V}^{}\right|_{}^2-\left| C_{\rm A}^{}\right|_{}^2=0$ and thus the dominant contribution to $a_\mu^{}$ is given by $g_{\rm FV}^{(2)}(y,Q_V^{})$ in Eq.~(\ref{eq:FV series}). In this case, one can conclude that $\Delta a_\mu^{\rm FV} >0$ for $Q_V\leq-5/9$ and $\Delta a_\mu^{\rm FV} <0$ for $Q_V\geq -0.433$.
\end{itemize}

%%%%%%%%%%%%%%%%%%%%%%%%%%%%%%%%%%%% Fig 4 %%%%%%%%%%%%%%%%%%%%%%%%%%%%%%%
%%%%%%%%%%%%%%%%%%%%%%%%%%%%%%%%%%%%%%%%%%%%%%%%%%%%%%%%%%%%%%%%%%%%%%%%%%

%%%%%%%%%%%%%%%%%%%%%%%%%%%%%%%%%%% Table 2 %%%%%%%%%%%%%%%%%%%%%%%%%%%%%%
\begin{table}[t!]
\renewcommand\arraystretch{2.2}
\centering
\begin{tabular}{c|c|c|c|c|c}
\hline \hline
{\bf Loop functions} & $y\to 0$ & $y\to \infty$& $y \to 1$ & $Q_{V}^{{\rm L}}$ & $Q_V^{{\rm U}}$ \\
\hline\hline
\multirow{2}{*}{$g_{\rm FV}^{(1)}\left(y,Q_S^{}\right)$}
 & \multirow{2}{*}{0} & $\infty$ (for $Q_V^{}\neq-1/2$) & \multirow{2}{*}{$2+5Q_{V}^{}$} &  \multirow{2}{*}{$\displaystyle -\frac{1}{2}$} & \multirow{2}{*}{ $-0.387$}\\
~&~& $0$ (for $Q_V^{}=-1/2$) &&\\
\hline 
$g_{\rm FV}^{(2)}\left(y,Q_S^{}\right)$ & $\displaystyle -\frac{8}{3}-6Q_V^{}$ & $\displaystyle -\frac{5}{3}-3Q_V^{}$ & $\displaystyle -\frac{13}{6}-5Q_V^{}$ & $\displaystyle -\frac{5}{9}$ & $-0.433$
\\
\hline\hline
\end{tabular}
\vspace{0.5cm}
\caption{Summary of the asymptotic values of $g_{\rm FV}^{(1)}(y,Q_V^{})$ and $g_{\rm FV}^{(2)}(y,Q_V^{})$ in the limit of $y \to 0$, $y\to \infty$ and $y\to 1$, as well as the corresponding critical values $Q_V^{\rm U}$ and $Q_V^{\rm L}$. Note that the upper-critical charges $Q_V^{\rm U}$ are irrational numbers and thus three significant digits are maintained.}
\label{table:FV}
\end{table}
\renewcommand\arraystretch{1}
%%%%%%%%%%%%%%%%%%%%%%%%%%%%%%%%%%%%%%%%%%%%%%%%%%%%%%%%%%%%%%%%%%%%%%%%%%

The loop functions of the FV-type are quite different from those of the FS-type. In particular, the divergence in the first-order loop function $g_{\rm FV}^{(1)}(y,Q_V^{})$ as $M_F^{}/M_V^{}\to \infty$ stems from the fact that the power of $y$ in the numerator of $g_{\rm FV}^{(1)}(y,Q_V^{})$ is higher than that in its denominator, while the power of $y$ in the numerator of $g_{\rm FS}^{(1)}(y,Q_S^{})$ is lower than that in the denominator.\footnote{If one calculates the Feynman amplitudes in Fig.~\ref{fig:FV} in the Feynman gauge, then the highest power of $y$ in the numerator will be reduced by two and there is no divergence in $g_{\rm FV}^{(1)}(y,Q_V^{})$~\cite{Athron2021}. However, there will be contributions from extra diagrams corresponding to the couplings among muon, $F$ and Goldstone bosons, leading to divergences as $M_F^{} \to \infty$.} However, as a physical observable, the muon anomalous magnetic moment should not blow up as the mass of $F$ approaches infinity. The only way for $a^{}_\mu$ to be finite in the limit of $M^{}_F \to \infty$ is to demand that the couplings between muon and $F$ should be further suppressed by the inverse powers of $M^{}_F$. Therefore, in the theory with a heavy fermion coupled to muon and a vector boson, we can draw the following model-independent constraint: \emph{If the heavy fermion is coupled simultaneously to both left- and right-handed muon and the vector boson, then the couplings should decrease at least as ${\cal O}\left(1/\sqrt{M_F^{}}\right)$ in the limit of $M^{}_F \to \infty$.} For an ultraviolet-complete model, this is indeed satisfied since the couplings are usually proportional to the mixing parameters between muon and $F$, which are suppressed by the heavy fermion mass.

\subsubsection{$m_{\mu}^{}, M_F^{} \ll M_V^{}$}
Then we assume that both muon and $F$ are much lighter than $V$, i.e., $m_{\mu}, M_F^{} \ll M_V^{}$.  To the order of ${\cal O}\left(m_{\mu}^2/M_V^2\right)$ one obtains
\begin{eqnarray}
\label{eq:FV appro2}
f_{\rm FV}^{}=-\frac{2}{3}\left(\frac{m_{\mu}}{M_V}\right)_{}^2\left[
\left(\left|C_{\rm V}^{}\right|_{}^2+ \left|C_{\rm A}^{}\right|_{}^2\right)\left(4+9Q_V^{}\right)-6\left(\left|C_{\rm V}^{}\right|_{}^2-\left|C_{\rm A}^{}\right|_{}^2\right)\left(1+2Q_V^{}\right)\frac{M_F^{}}{m_\mu^{}}
\right]
\; ,
%     (27)
\end{eqnarray}
where the mass ratio $M_F^{}/m_{\mu}^{}$ can take any positive values. In the limit of $M_F^{}/m_\mu^{}\to 0$, which is
valid for $F$ being neutrinos in the SM, the loop function in Eq.~(\ref{eq:FV appro2}) is reduced to
\begin{eqnarray}
\label{eq:FV appro2p}
f_{\rm FV}^{}=-\frac{2}{3}\left(\frac{m_{\mu}}{M_V}\right)_{}^2\left(\left|C_{\rm V}^{}\right|_{}^2+ \left|C_{\rm A}^{}\right|_{}^2\right)\left(4+9Q_V^{}\right)\;,
%     (28)
\end{eqnarray}
where the term proportional to $\left(\left| C_{\rm V}^{}\right|_{}^2-\left| C_{\rm A}^{}\right|_{}^2\right)$ vanishes. As indicated by Eq.~(\ref{eq:FV appro2p}), in this case, the contribution to $a_\mu^{}$ is positive only for $Q_V^{}<-4/9$.

Moreover, if $m_\mu^{}\ll M_F^{}$ holds, then Eq.~(\ref{eq:FV appro2}) can be greatly simplified to
\begin{eqnarray}
\label{eq:FV appro2pp}
f_{\rm FV}^{}=4\,\frac{m_\mu M_F}{M_V^2}\left(\left|C_{\rm V}^{}\right|_{}^2-\left|C_{\rm A}^{}\right|_{}^2\right)\left(1+2Q_V^{}\right)\;,
\end{eqnarray}
which can also be obtained by setting $y\equiv M_F^{}/M_V^{}\ll 1$ in the function $g_{\rm FV}^{(1)}(y,Q_V^{})$ in Eq.~(\ref{eq:gFV1}).

\subsubsection{$m_{\mu}^{}, M_V^{} \ll M_F^{}$}
Finally we consider the scenario where both muon and $V$ are much lighter than $F$, i.e., $m_{\mu}^{}, M_V^{} \ll M_F^{}$. To the order of ${\cal O}\left(m_{\mu}^2/M_F^2\right)$, we have
\begin{eqnarray}
\label{eq:FV appro3}
f_{\rm FV}^{}=\left(\left|C_{\rm V}\right|_{}^2-\left|C_{\rm A}\right|_{}^2\right)\left(1+2Q_V^{}\right)\frac{m_{\mu} M_F}{M_V^2}-\frac{1}{3}\left(\left|C_{\rm V}\right|_{}^2+\left|C_{\rm A}\right|_{}^2\right)\left(5+9Q_V^{}\right)\left(\frac{m_{\mu}}{M_V}\right)_{}^2\;,
%     (29)
\end{eqnarray}
which can also be obtained by taking the limit of $y = M_F^{}/M_V^{}\to \infty$ in Eqs.~(\ref{eq:gFV1}) and (\ref{eq:gFV2}). Two interesting observations can be made.

First, as has been emphasized in the last paragraph of Sec.~\ref{subsec:mmu<< MF,MV}, for the theory with a heavy fermion coupled simultaneously to both left- and right-handed muon (i.e., $\left|C_{\rm V}\right|_{}^2-\left|C_{\rm A}\right|_{}^2\neq 0$), the first-order loop function, which scales as ${\cal O}\left(M_F^{} m_{\mu}^{}/M_V^2\right)$, will be divergent for $M^{}_F \to \infty$. Hence a finite contribution to $a_\mu^{}$ requires $C_{\rm V}^{}$ and $C_{\rm A}^{}$ 
to decrease no more slowly than ${\cal O}\left(1/\sqrt{M_F^{}}\right)$.

Second, the right-hand side of Eq.~(\ref{eq:FV appro3}) becomes singular as $M_V^{}$ approaches zero. In order to track this singularity one can take $M_V^{}=0$ from the very beginning in Eqs.~(\ref{eq:FVa amp}) and (\ref{eq:FVb amp}).\footnote{Note that when taking $M_V^{}=0$, we have to modify the propagator of the vector boson in Eq.~(\ref{eq:FVb amp}) to $-{\rm i}\eta_{\rho\sigma}^{}/k^2_{}$.} After doing so, one can find that the contribution from Fig.~\ref{fig:FV}(b) is finite while that from Fig.~\ref{fig:FV}(a) is infinite. This result can be understood by recalling the famous Weinberg-Witten theorem~\cite{WW1980}: \emph{For the theory with a Lorentz-covariant conserved current, massless particles of spin $j>1/2$ with a nonvanishing charge, corresponding to the conserved current, cannot exist.} In our case, the particles running in the loop are coupled to the external photon via the electromagnetic interaction, and the corresponding electromagnetic current is of course Lorentz-covariant and conserved. The Weinberg-Witten theorem excludes the existence of a massless vector boson with a nonvanishing electric charge. In other words, the vector boson in the Feynman diagram in Fig.~\ref{fig:FV}(a) cannot be massless, implying the absence of the aforementioned singularity.

\subsection{The FSV-type}
\label{sec:F+S+V}
%%%%%%%%%%%%%%%%%%%%%%%%%%%%%%% Fig. 5 %%%%%%%%%%%%%%%%%%%%%%%%%%%%%%%%
\begin{figure}[t!]
	\centering		\includegraphics[width=17.5cm]{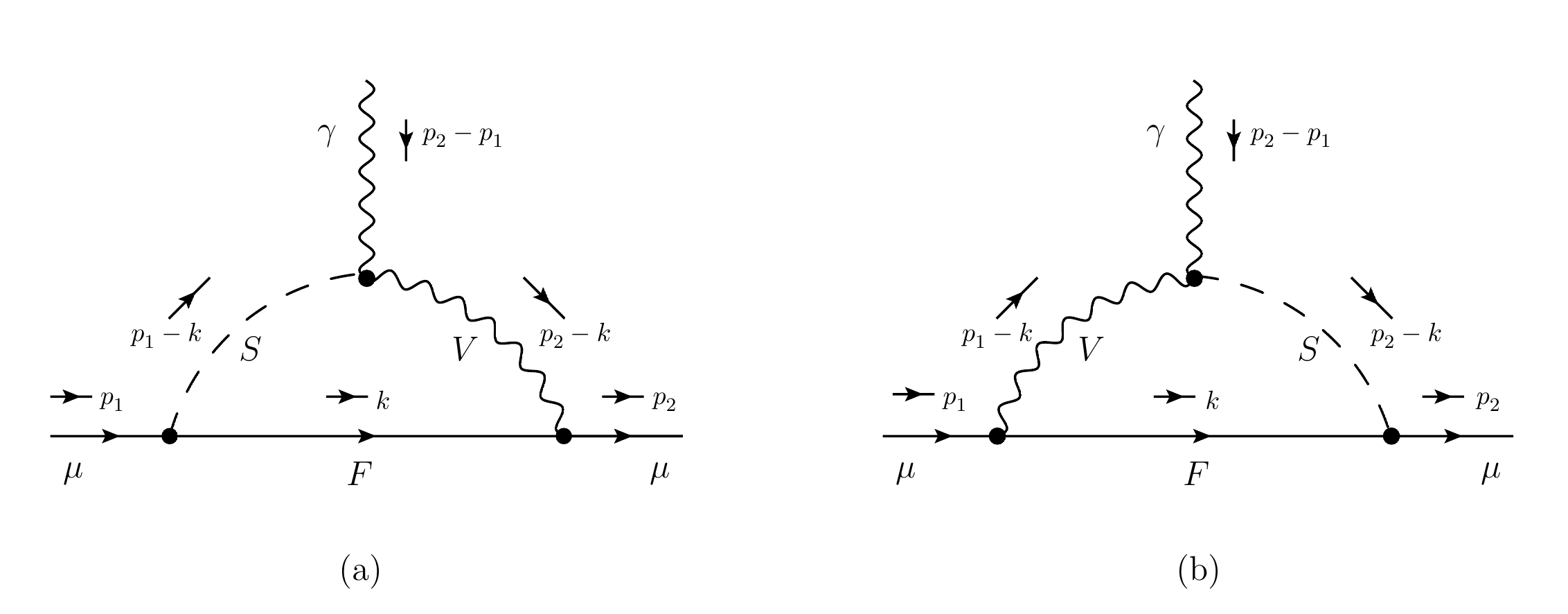}
		\vspace{-0.3cm}
	\caption{The Feynman diagrams for the FSV-type contributions to the muon anomalous magnetic moment $a^{}_\mu$, where the particle four-momenta have been explicitly specified and represented by arrows along the internal and external lines.}
	\label{fig:FSV}
\end{figure}
%%%%%%%%%%%%%%%%%%%%%%%%%%%%%%%%%%%%%%%%%%%%%%%%%%%%%%%%%%%%%%%%%%%%%%%
In this subsection, we study the scenario where there is a fermion $F$, a scalar $S$ and a vector boson $V$ in the loops that contribute to $a_\mu^{}$. The relevant Feynman diagrams are shown in Fig.~(\ref{fig:FSV}). These FSV-type diagrams usually appear in the calculations in the $R_{\xi}^{}$ gauge, for which there are extra contributions from Goldstone bosons. However, since we perform all the calculation in the unitary gauge, there are no such diagrams in the SM. In the extensions of the SM, such diagrams indeed exist and $F$, $S$ and $V$ can all be identified with physical particles. A typical example is the type-II seesaw model, which will be discussed in Sec.~\ref{subsec:type-II seesaw}. 

As in the previous cases, we write down the general Lagrangian for the relevant particles contributing to $a_\mu^{}$, i.e.,\footnote{Here we only include the renormalizable interactions. The cases of the axion and axion-like particles~\cite{Ge2021,Keung2021,Brdar2021,Buen-Abad2021,Bhattacharya2021,Barr1990,Chang2000,Marciano2016,Bauer2019,Cornella2019}, which contribute to the muon $g-2$ via the dimension-five non-renormalizable effective operator, are not included.}
\begin{eqnarray}
\label{eq:FSV langrangian}
{\cal L}_{\rm int}^{\rm FSV}=\left[\overline{F}\left( C_{\rm S}^{}+ C_{\rm P}^{}\gamma_5^{}\right)\mu S_{}^{\dagger}+\overline{F}\left(C_{\rm V}^{}\gamma_{}^{\mu}+C_{\rm A}^{}\gamma_{}^{\mu}\gamma_5^{} \right)\mu V_{\mu}^{\dagger} +{\rm h.c.}\right]+\mu_0^{}e A^{\mu}_{}\left(V_{\mu}^{\dagger}S+V_{\mu}S_{}^{\dagger}\right)\;,
%     (30)
\end{eqnarray}
where $\mu_0^{}$ stands for a free parameter of mass dimension and will be fixed in a concrete model. 

The amplitudes corresponding to the Feynman diagrams in Fig.~\ref{fig:FSV}(a) and Fig.~\ref{fig:FSV}(b) are given by
\begin{eqnarray}
\label{eq:FSVa amp}
\overline{u}(p_2^{})\left(-{\rm i} e \Gamma_{\rm FSV}^{\mu,({\rm a})}\right)u(p_1^{})&=&{\rm i}\mu_0^{}e \int\frac{{\rm d}^4 k}{\left(2\pi\right)^4}\left\{\overline{u}(p_2^{}){\rm i}\left(C_{\rm V}^{*}\gamma_{}^{\alpha}+C_{\rm A}^{*}\gamma_{}^{\alpha}\gamma_5^{}\right)\frac{{\rm i}}{\slashed{k}-M_F}\right.\nonumber\\
&&\left. \times \left(C_{\rm S}+C_{\rm P}\gamma_5^{}\right)u(p_1)\frac{\rm i}{\left(p_1-k\right)^2-M_S^2}\frac{-{\rm i\eta^{\mu \beta}}}{\left(p_2-k\right)^2-M_V^2}\right.\nonumber\\
&&\left. \times \left[\eta_{\alpha \beta}^{}-\frac{\left(p_2-k\right)_{\alpha}\left(p_2-k\right)_{\beta}}{M_V^2} \right]
\right\}\;,
%     (31)
\end{eqnarray}
and
\begin{eqnarray}
\label{eq:FSVb amp}
\overline{u}(p_2^{})\left(-{\rm i} e \Gamma_{\rm FSV}^{\mu,({\rm b})}\right)u(p_1^{})&=&
{\rm i}\mu_0^{}e \int\frac{{\rm d}^4 k}{\left(2\pi\right)^4}\left\{\overline{u}(p_2^{}){\rm i}\left(C_{\rm S}^{*}-C_{\rm P}^{*}\gamma_5^{}\right)\frac{{\rm i}}{\slashed{k}-M_F}\right.\nonumber\\
&&\left. \times \left(C_{\rm V}\gamma_{}^{\alpha}+C_{\rm A}\gamma_{}^{\alpha}\gamma_5^{}\right)u(p_1)\frac{\rm i}{\left(p_2-k\right)^2-M_S^2}\frac{-{\rm i\eta^{\mu \beta}}}{\left(p_1-k\right)^2-M_V^2}\right.\nonumber\\
&&\left. \times \left[\eta_{\alpha \beta}^{}-\frac{\left(p_1-k\right)_{\alpha}\left(p_1-k\right)_{\beta}}{M_V^2} \right]
\right\}\; .
%     (32)
\end{eqnarray}
From these amplitudes, one can extract the FSV-type contributions to $a_\mu^{}$ as follows
\begin{eqnarray}
\label{eq:FSV amu}
a_{\mu}^{\rm FSV}=\frac{1}{16 \pi^2}\left(\frac{\mu_0}{M_V}\right)f_{\rm FSV}^{}\left(\frac{m_\mu}{M_V},\frac{m_\mu}{M_S},\frac{m_\mu}{M_F},C_{\rm S}^{},C_{\rm P}^{},C_{\rm V}^{},C_{\rm A}^{}\right)\;,
%     (33)
\end{eqnarray}
where the loop function can be decomposed into two parts\footnote{In the special case, where both $S$ and $V$ are only coupled to either left-handed or right-handed muon, then the first part vanishes. If $S$ is only coupled to left-handed (or right-handed) muon and $V$ to right-handed (or left-handed) muon, then the second part vanishes.}
\begin{eqnarray}
\label{eq:FSV}
f_{\rm FSV}^{}\left(x,y,z,C_{\rm S}^{},C_{\rm P}^{},C_{\rm V}^{},C_{\rm A}^{}\right)&=& +\left[ {\rm Re} \left(C_{\rm A}^{*}C_{\rm P}^{}\right) - {\rm Re} \left(C_{\rm S}^{*}C_{\rm V}^{}\right) \right] f_{\rm FSV}^{-}\left(x,y,z\right)\nonumber\\
&&+\left[ {\rm Re} \left(C_{\rm A}^{*}C_{\rm P}^{}\right) + {\rm Re} \left(C_{\rm S}^{*}C_{\rm V}^{}\right) \right]f_{\rm FSV}^{+}\left(x,y,z\right)\;,
%     (34)
\end{eqnarray}
with
\begin{eqnarray}
\label{eq:FSV-}
f_{\rm FSV}^{-}\left(x,y,z\right)&=&
\frac{1}{3 x^3 y^4 z^6 \left(x^2-y^2\right)^2}\left\{
-2x_{}^2y_{}^2z_{}^6\left(x_{}^2-y_{}^2\right)\left[x_{}^4+y_{}^4+x_{}^2y_{}^2\left(2y_{}^2-x_{}^2-1\right)
\right]\right.\nonumber\\
&& \left. +x_{}^4y_{}^4z_{}^4\left(x_{}^4-y_{}^4\right)+y_{}^6\ln\left(\frac{x}{y}\right)\left[2y_{}^2 z_{}^6+x_{}^2z_{}^4\left(3\left(3x_{}^2-y_{}^2\right)\left(1+z_{}^2\right)-4z_{}^2
\right) \right.\right.\nonumber\\
&&\left.\left. +x_{}^6\left(1-z_{}^2\right)\left(\left(x_{}^2+y_{}^2\right)\left(1-z_{}^2\right)_{}^2-6z_{}^2\left(1+z_{}^2\right)\right)\right]+z_{}^4\left(x_{}^2-y_{}^2\right)_{}^3\ln\left(\frac{y}{z}\right)\right.\nonumber\\
&&\left. \times\left[-2\left(x_{}^2+y_{}^2\right)z_{}^2+3x_{}^2y_{}^2\left(1+z_{}^2\right)\right]+x_{}^6y_{}^2z_{}^2\Lambda\left(1,\frac{1}{y},\frac{1}{z}\right)
\left[2z_{}^4\left(2y_{}^2-x_{}^2\right)\right.\right.\nonumber\\
&&\left.\left. +y_{}^4\left(x_{}^2+y_{}^2\right)\left(1-z_{}^2\right)_{}^2+y_{}^2z_{}^2\left(x_{}^2-5y_{}^2\right)\left(1+z_{}^2\right)
\right]-y_{}^6x_{}^2z_{}^2\Lambda\left(1,\frac{1}{x},\frac{1}{z}\right)\right.\nonumber\\
&&\left. \times \left[2z_{}^4\left(2x_{}^2-y_{}^2\right)+x_{}^4\left(y_{}^2+x_{}^2\right)\left(1-z_{}^2\right)_{}^2+x_{}^2z_{}^2\left(y_{}^2-5x_{}^2\right)\left(1+z_{}^2\right)\right]
\right\}
\;, \quad
%     (35)
\end{eqnarray}
and
\begin{eqnarray}
\label{eq:FSV+}
f_{\rm FSV}^{+}\left(x,y,z\right)&=&
\frac{1}{3x^3y^4z^7\left(x^2-y^2\right)}\left\{-x_{}^2y_{}^2z_{}^4\left(x_{}^2-y_{}^2\right)\left[2y_{}^2\left(z_{}^2-2x_{}^2\right)+x_{}^2z_{}^2\left(2+y_{}^2\right)
\right]\right.\nonumber\\
&&\left. +2y_{}^6\ln\left(\frac{x}{y}\right)\left[z_{}^4\left(z_{}^2-3x_{}^2\right)+x_{}^4\left(1-z_{}^2\right) \left(3z_{}^2-x_{}^2\left(1-z_{}^2\right)_{}^2\right)\right]-2z_{}^2\right.\nonumber\\
&&\left. \times\left(x_{}^2-y_{}^2\right)\ln \left(\frac{y}{z}\right)
\left[3x_{}^2y_{}^2\left(x_{}^2y_{}^2\left(1-z_{}^2\right)-z_{}^2\left(x_{}^2+y_{}^2\right)\right)+z_{}^4\left(x_{}^4+x_{}^2y_{}^2+y_{}^4\right)\right]\right.\nonumber\\
&&\left. +2x_{}^2y_{}^6z_{}^2\Lambda\left(1,\frac{1}{x},\frac{1}{z}\right)
\left[z_{}^2\left(z_{}^2-x_{}^2\right)-x_{}^2\left(1-z_{}^2\right)\left(z_{}^2-x_{}^2+x_{}^2z_{}^2\right)\right]\right.\nonumber\\
&&\left. -2y_{}^2x_{}^6z_{}^2\Lambda\left(1,\frac{1}{y},\frac{1}{z}\right)
\left[z_{}^2\left(z_{}^2-y_{}^2\right)-y_{}^2\left(1-z_{}^2\right)\left(z_{}^2-y_{}^2+y_{}^2z_{}^2\right)\right]
\right\}
\;.
%     (36)
\end{eqnarray}
Notice that the last term in $f_{\rm FSV}^{-}\left(f_{\rm FSV}^{+}\right)$ can be obtained by simply exchanging between $x$ and $y$ in the next-to-last term in $f_{\rm FSV}^{-}\left(f_{\rm FSV}^{+}\right)$ up to a minus sign.

Although the general expressions in Eqs.~(\ref{eq:FSV-}) and (\ref{eq:FSV+}) are more complicated than those in the previous cases, we have included these exact results for completeness. In the SM extensions, these formulae may find some practical applications. In fact, the explicit expressions can be greatly simplified under the conditions of strong mass hierarchies. For instance, we assume that muon is much lighter than $V$ and $S$, i.e., $m_{\mu}\ll M_V^{}, M_S^{}$. Hence, to the order of ${\cal O}\left(m_{\mu}^3/M_V^3\right)$, one has
\begin{eqnarray}
\label{eq:FSV appro}
f_{\rm FSV}^{}&=&  +\left\{\left[{\rm Re} \left(C_{\rm A}^{*}C_{\rm P}^{}\right) - {\rm Re} \left(C_{\rm S}^{*}C_{\rm V}^{}\right)\right]g_{\rm FSV}^{(1)}\left(\frac{M_V}{M_S}\right)\right\} \left(\frac{m_{\mu}}{M_V}\right) \nonumber\\
&& +\left\{\left[{\rm Re} \left(C_{\rm A}^{*}C_{\rm P}^{}\right) - {\rm Re} \left(C_{\rm S}^{*}C_{\rm V}^{}\right)\right]g_{\rm FSV}^{(3-)}\left(\frac{M_V}{M_S},\frac{M_F}{m_{\mu}}\right)\right.\nonumber\\
&&\left. +\left[{\rm Re} \left(C_{\rm A}^{*}C_{\rm P}^{}\right) + {\rm Re} \left(C_{\rm S}^{*}C_{\rm V}^{}\right)\right]g_{\rm FSV}^{(3+)}\left(\frac{M_V}{M_S},\frac{M_F}{m_{\mu}}\right)
\right\}\left(\frac{m_{\mu}}{M_V}\right)_{}^3\;,
%     (37)
\end{eqnarray}
where
\begin{eqnarray}
\label{eq:gFSV1}
g_{\rm FSV}^{(1)}\left(x\right)&=&\frac{-2x^2}{(x^2-1)^2}\left(1-x^2_{}+2\ln x\right)\;,\\
%     (38)
\label{eq:gFSV3-}
g_{\rm FSV}^{(3-)}\left(x,y\right)&=&\frac{-2x^2}{3(x^2-1)^2}\left[\left(1+x_{}^2\right)\ln x+1-x_{}^2 \right]\left(3y_{}^2-1\right)\;,\\
%     (39)
\label{eq:gFSV3+}
g_{\rm FSV}^{(3+)}\left(x,y\right)&=&\frac{4x^2 y}{3(x^2-1)}\ln x\;.
%     (40)
\end{eqnarray}
Note that the loop functions in Eqs.~(\ref{eq:gFSV1})-(\ref{eq:gFSV3+}) are all continuous at $x=1$. More explicitly, we have the finite limits
\begin{eqnarray*}
\lim_{x\to 1} g_{\rm FSV}^{(1)}\left(x\right)=1\;,\quad
\lim_{x\to 1} g_{\rm FSV}^{(3-)}\left(x,y\right)=0\;,\quad
\lim_{x\to 1} g_{\rm FSV}^{(3+)}\left(x,y\right)=\frac{2y}{3}\;,
\end{eqnarray*}
indicating the absence of any singularities at $M_V^{}=M_S^{}$. In general, it is impossible to reach a simple conclusion on the sign of the loop functions or that of $a^{\rm FSV}_\mu$. The situation is even worse for the exact loop functions in Eqs.~(\ref{eq:FSV-}) and (\ref{eq:FSV+}). However, if ${\rm Re}\left(C^*_{\rm A} C^{}_{\rm P}\right) = {\rm Re}\left(C^*_{\rm S} C^{}_{\rm V}\right)$ holds, then we can see from Eq.~(\ref{eq:FSV appro}) that only the loop function $g^{(3+)}_{\rm FSV}(M^{}_V/M^{}_S, M^{}_F/m^{}_\mu)$ is involved. In this special case, one can immediately verify that $g^{(3+)}_{\rm FSV}(M^{}_V/M^{}_S, M^{}_F/m^{}_\mu)$ in Eq.~(\ref{eq:gFSV3+}) will always be positive for arbitrary mass ratios $M^{}_V/M^{}_S$ and $M^{}_F/m^{}_\mu$. Thus the sign of $a_\mu^{}$ is just determined by the sign of $\mu_0^{}\left[{\rm Re} \left(C_{\rm A}^{*}C_{\rm P}^{}\right) + {\rm Re} \left(C_{\rm S}^{*}C_{\rm V}^{}\right)\right]$.

\section{Practical Applications}
\label{sec:examples}
In this section, we take some concrete models, which have been extensively discussed in the literature, as examples to illustrate how our general formulae derived in Sec.~\ref{sec:loop} can be practically applied to the calculations of muon $g-2$. On the other hand, the latest measurement of the fine structure constant from the recoil velocity of rubidium atom brings about a 1.6 $\sigma$ discrepancy\footnote{Notice that an earlier measurement using the caesium atom leads to a 2.4 $\sigma$ discrepancy with an opposite sign~\cite{Parker2018}.} with the theoretical prediction for the electron anomalous magnetic moment~\cite{Morel2020}. We emphasize that our general formulae derived in Sec.~\ref{sec:loop} can be directly used to calculate the one-loop contributions to electron $g-2$ by only replacing the mass and couplings of muon with those of electron. However, in the remaining part of this section, we shall focus on the one-loop contributions to muon anomalous magnetic moment for clarity.

In the subsequent discussions, we first apply our general formulae to the SM and rederive the well-known one-loop QED and electroweak (EW) contributions to muon $g-2$. Then, we investigate several new-physics models, including the type-I seesaw model, the type-II seesaw model, the $Z^{\prime}_{}$ model and the scalar leptoquark model. Special attention will be paid to a quick judgment about the sign of $\Delta a_\mu^{}$ in concrete models.

\subsection{SM contributions}
%%%%%%%%%%%%%%%%%%%%%%%%%%%%%%% Fig. 6 %%%%%%%%%%%%%%%%%%%%%%%%%%%%%%%%
\begin{figure}[t!]
	\centering		\includegraphics[width=18cm]{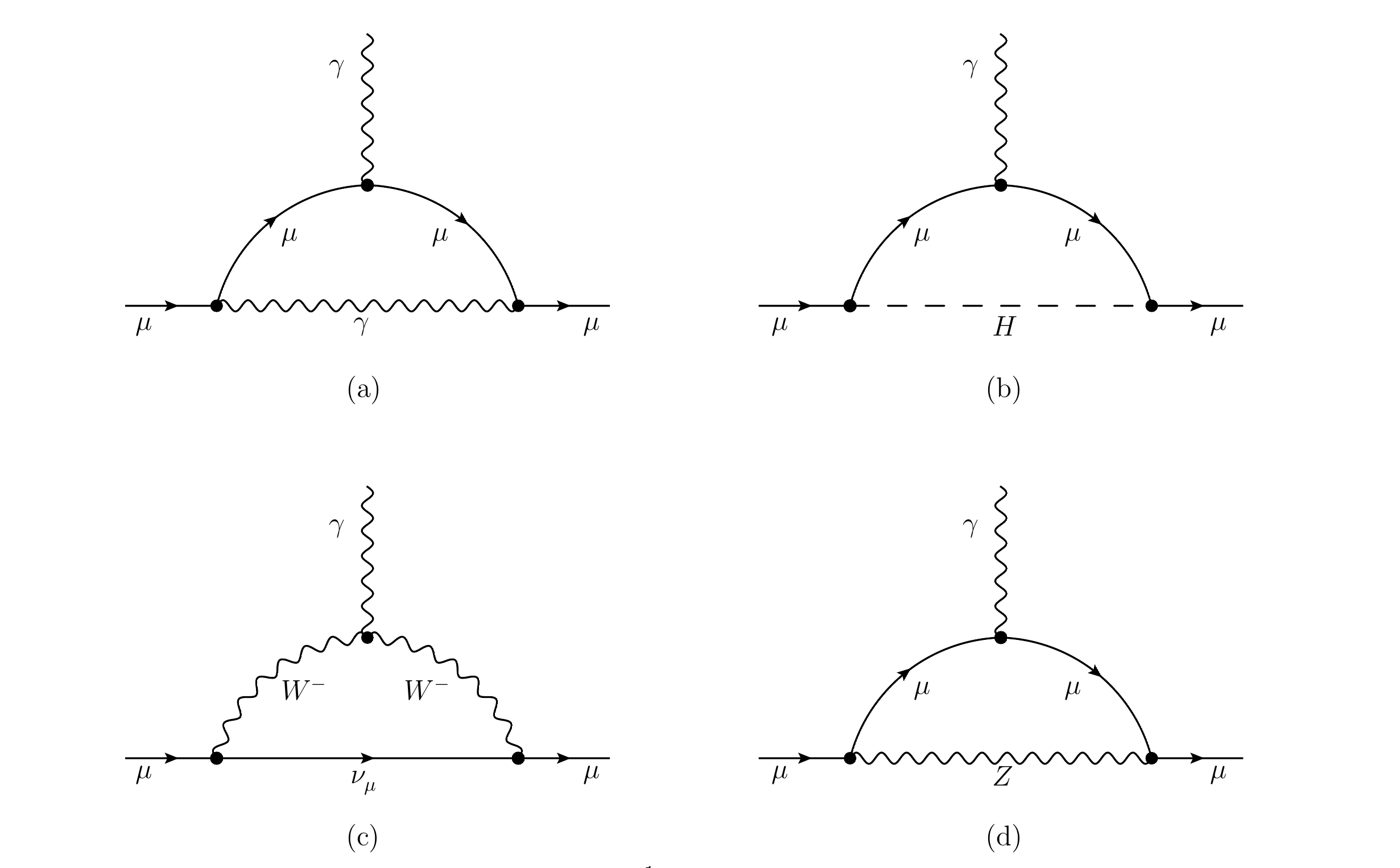}
		\vspace{-0.3cm}
	\caption{The Feynman diagrams for the one-loop QED and EW corrections to the muon $g-2$ in the SM: (a) QED; (b) SM Higgs boson; (c) SM charged gauge boson $W_{}^{-}$; (d) SM neutral gauge boson $Z$.}
	\label{fig:SM}
\end{figure}
%%%%%%%%%%%%%%%%%%%%%%%%%%%%%%%%%%%%%%%%%%%%%%%%%%%%%%%%%%%%%%%%%%%%%%%
Since Julian Schwinger first calculated the one-loop QED correction to the anomalous magnetic moment of electron in 1948~\cite{Schwinger1948}, great efforts have been made to the precise calculation of higher-order corrections to the anomalous magnetic moment of charged leptons. See, for example, Ref.~\cite{Aoyama2020}, for a recent review. The one-loop EW corrections to muon $g-2$ were calculated more than fifty years ago~\cite{Brodsky1966,Burnett1967,Jackiw1972,Bars1972,Fujikawa1972,Altarelli1972,Bardeen1972}. In this subsection, we reproduce the well-known results of one-loop QED and EW contributions to $a_\mu^{}$ by using our general formulae derived in the last section.

\begin{itemize}
\item \emph {QED.} The one-loop QED contribution to the muon anomalous magnetic moment belongs to the FV-type, corresponding to Fig.~\ref{fig:SM}(a). Setting $Q_V^{}=0$, $M_F^{}=m_{\mu}^{},M_{V}^{}=0$, $C_{\rm V}^{}=e$ and $C_{\rm A}^{}=0$ in Eqs.~(\ref{eq:FV amu})-(\ref{eq:FV+}), one obtains
\begin{eqnarray}
a_{\mu}^{{\rm QED}}=\frac{1}{16\pi^2}{\lim_{x \to \infty}}f_{\rm FV}^{}\left(x,x,e,0,0\right)=\frac{\alpha}{2\pi}\;,
%     (41)
\end{eqnarray}
where $\alpha=e_{}^2/(4\pi)$ is the fine structure constant. Note that the massless photon in the loop is not forbidden by Weinberg-Witten theorem since it is neutral and thus not coupled to the Lorentz-covariant conserved electromagnetic current. As a consequence, the contribution from Fig.~\ref{fig:FV}(b) is actually finite in the limit of $M_V^{} \to 0$.

\item \emph {Higgs boson.} The SM Higgs boson contributes to the muon anomalous magnetic moment at one-loop level via Fig.~\ref{fig:SM}(b), belonging to the FS-type. Since the Higgs boson is much heavier than muon, it is safe to use the approximate expression in Eq.~(\ref{eq:FS appro2}). Taking $Q_S^{}=0$, $M_F^{}=m_\mu^{}$, $C_{\rm S}^{}=-\sqrt{2\lambda}\,m_{\mu}^{}/M_{H}$ and $C_{\rm P}^{}=0$, we arrive at
\begin{eqnarray}
\label{eq:SM Higgs contribution}
a_{\mu}^{\rm Higgs}=-\frac{\lambda}{24\pi^2}\left[7+12\ln\left(\frac{m_{\mu}^{}}{M_H}\right)\right]\left(\frac{m_{\mu}}{M_H}\right)_{}^4\;,
%     (42)
\end{eqnarray}
where $\lambda$ and $M_H^{}$ are the quartic self-coupling and Higgs-boson mass, respectively. One can check that the result in Eq.~(\ref{eq:SM Higgs contribution}) is equivalent to that of the Higgs contribution in Ref.~\cite{Leveille1977} after completing the Feynman integral, which is proportional to the integral in Eq.~(\ref{eq:integrand}), and retaining the leading-order term of $m_\mu^{}/M_H^{}$. However, our method is much more convenient since the integral in Eq.~(\ref{eq:integrand}) is rather tedious to perform analytically without imposing any conditions on mass hierarchies. 

\item \emph {$W_{}^{}$ and $Z$ bosons.} The $W$ and $Z$ bosons contribute to the muon $g-2$ at the one-loop level via Fig.~\ref{fig:SM}(c) and Fig.~\ref{fig:SM}(d), belonging to the FV-type. Substituting $Q_V^{}=1$, $C_{\rm V}^{}=-C_{\rm A}^{}=g/(2\sqrt{2})$ into Eq.~(\ref{eq:FV appro2p}) and $Q_V^{}=0$, $C_{\rm V}^{}=g\left(-1+4\sin_{}^2\theta_{\rm W}^{}\right)/(4\cos\theta_{\rm W}^{})$, $C_{\rm A}^{}=g/(4\cos\theta_{\rm W}^{})$, $M^{}_F/m^{}_\mu = 1$ into Eq.~(\ref{eq:FV appro2}), respectively, we get
\begin{eqnarray}
\label{eq:SM W contribution}
a_{\mu}^{W}&=&\frac{5 G_{\rm F}^{}m_{\mu}^2}{12\sqrt{2}\pi^2}\;,\\
%     (43)
\label{eq:SM Z contribution}
a_{\mu}^{Z}&=&-\frac{G_{\rm F}m_{\mu}^2}{24\sqrt{2}\pi^2}\left[5-\left(1-4\sin_{}^2\theta_W^{}\right)_{}^2\right]\;,
%     (44)
\end{eqnarray}
where $G_{\rm F}^{}$ is the Fermi constant, $g$ is the ${\rm SU}(2)$ gauge coupling and $\theta_{\rm W}^{}$ is the Weinberg angle. Note that the positive sign of $a_{\mu}^{W}$ can be simply understood via the factor in Eq.~(\ref{eq:FV appro2p}), i.e., $-(2/3) \times (4+9Q_{W^-}^{})=+10/3>0$. As the Higgs contribution in Eq.~(\ref{eq:SM Higgs contribution}) is highly suppressed by the Yukawa coupling of muon, the total one-loop EW contributions to $a_\mu^{}$ turn out to be
\begin{eqnarray}
a_{\mu}^{\rm EW}\approx a_{\mu}^{W}+a_{\mu}^Z=\frac{G_{\rm F}m_\mu^2}{24\sqrt{2}\pi^2}\left[5+\left(1-4\sin_{}^2\theta_W^{} \right)_{}^2\right]\;,
%     (45)
\end{eqnarray}
which perfectly reproduces the well-known result in the literature~\cite{Aoyama2020,Marciano2002,Gnendiger2013}.
\end{itemize}

Thus far we have rederived the one-loop QED and EW corrections to $a_\mu^{}$. Although the results are not new at all, we have demonstrate that our formulae are advantageous in avoiding any tedious integrals stemming from the Feynman parametrization. More examples about new physics models contributing to muon $g-2$ will be discussed below.

\subsection{Type-I seesaw model}
\label{subsec:type-I seesaw}
%%%%%%%%%%%%%%%%%%%%%%%%%%%%%%% Fig. 7 %%%%%%%%%%%%%%%%%%%%%%%%%%%%%%%%
\begin{figure}[t!]
	\centering		\includegraphics[width=9cm]{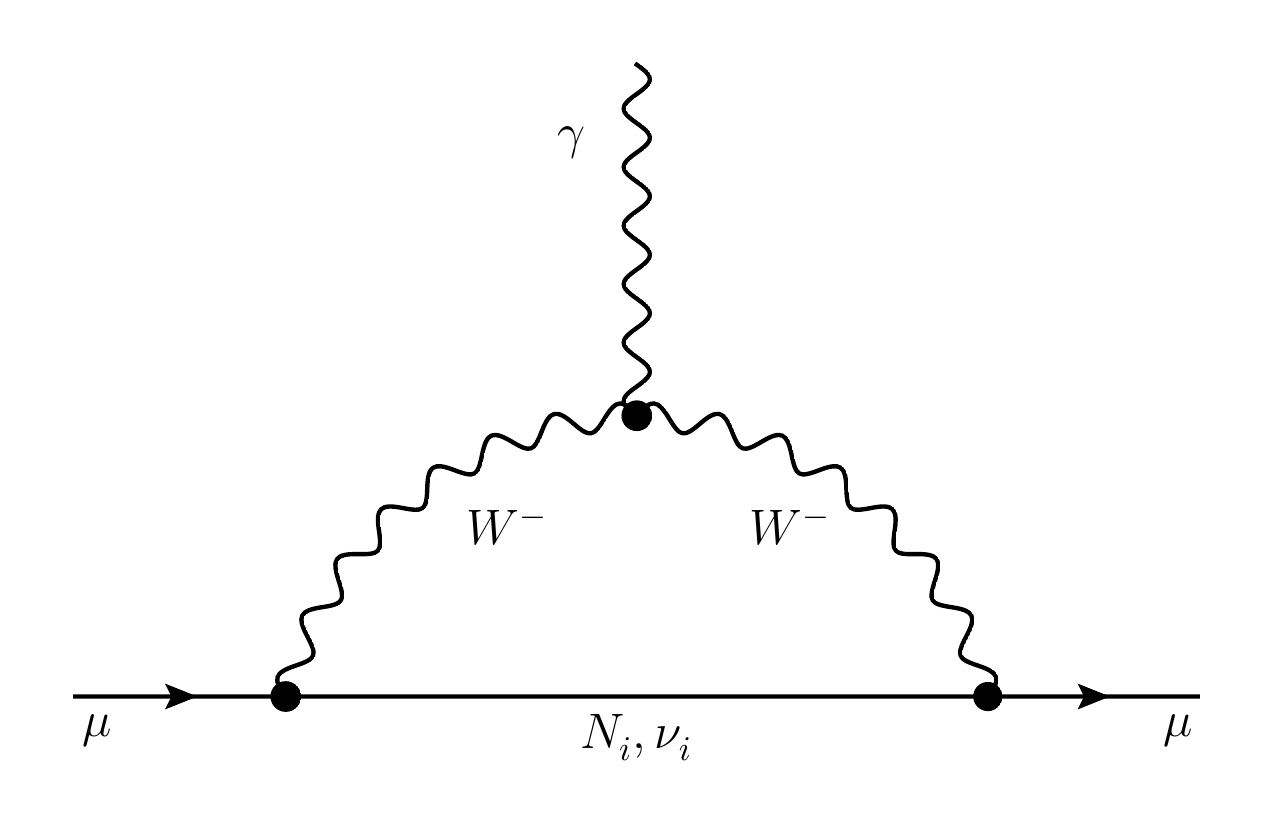}
		\vspace{-0.3cm}
	\caption{The Feynman diagram for the contributions from heavy and light Majorana neutrinos to the muon anomalous magnetic moment in the type-I seesaw model.}
	\label{fig:Type-I seesaw}
\end{figure}
%%%%%%%%%%%%%%%%%%%%%%%%%%%%%%%%%%%%%%%%%%%%%%%%%%%%%%%%%%%%%%%%%%%%%%%
The type-I seesaw model~\cite{Minkowski1977,Yanagida1980,GellMann1980,Glashow1979,Mohapatra1979} extends the SM by adding three right-handed neutrinos $N_{\rm R}^{}$, which are the SM gauge singlets and possess a large Majorana mass term. In this model, tiny Majorana masses of three active neutrinos can be generated via the seesaw mechanism on the one hand; the CP-violating and out-of-equilibrium decays of heavy Majorana neutrinos in the early Universe could explain the cosmological matter-antimatter asymmetry via the leptogenesis mechanism on the other hand. The gauge-invariant Lagrangian of the type-I seesaw model reads
\begin{eqnarray}
{\cal L}_{{\rm Type-I}}={\cal L}_{\rm SM}- \left(\overline{L}Y_{\nu}^{}\tilde{H}N_{\rm R} + \frac{1}{2}\overline{N_{\rm R}^{\rm C}}M_{\rm R}^{}N_{\rm R}^{}+{\rm h.c.} \right)\;,
%     (46)
\end{eqnarray}
where $L=\left(\nu_{\rm L}^{},l_{\rm L}^{}\right)_{}^{\rm T}$ and $\tilde{H}={\rm i}\sigma_2^{}H_{}^{*}$ with $H=\left(\varphi_{}^+,\varphi_{}^0\right)_{}^{\rm T}$ are the left-handed lepton doublet and the Higgs doublet, $Y_\nu^{}$ is the Dirac neutrino Yukawa coupling matrix while $M_{\rm R}^{}$ is the Majorana mass matrix for right-handed neutrino singlets. After the SM gauge symmetry is spontaneously broken down, the Dirac mass matrix linking the left-handed and right-handed neutrinos is given by $M_{\rm D}=Y_{\nu}^{}v/\sqrt{2}$ with $v \approx 246~{\rm GeV}$ the vacuum expectation value (VEV) of Higgs field. The overall $6\times 6$ neutrino mass matrix can be diagonalized by a $6\times 6$ unitary matrix via
\begin{eqnarray}
\left(
\begin{matrix}
{\cal V}& {\cal R}\\
{\cal S}& {\cal U}
\end{matrix}
\right)_{}^{\dagger}
\left(
\begin{matrix}
0&M_{\rm D}^{}\\
M_{\rm D}^{\rm T}&M_{\rm R}^{}
\end{matrix}
\right)
\left(
\begin{matrix}
{\cal V}& {\cal R}\\
{\cal S}& {\cal U}
\end{matrix}
\right)_{}^{*}=
\left(
\begin{matrix}
\widehat{M}_{\nu}^{}&0\\
0&\widehat{M}_{\rm R}^{}
\end{matrix}
\right)\;,
%     (47)
\end{eqnarray}
where $\widehat{M}_{\nu}={\rm diag}(m_1^{},m_2^{},m_3^{})$ and $\widehat{M}_{\rm R} = {\rm diag}(M_1^{},M_2^{},M_3^{})$ with $m_i^{} (M_i^{})$ (for $i=1,2,3$) being light (heavy) neutrino masses, and ${\cal V}$, ${\cal R}$, ${\cal S}$, ${\cal U}$ are all $3\times 3$  matrices satisfying unitarity conditions ${\cal V}{\cal V}_{}^{\dagger}+{\cal R}{\cal R}_{}^{\dagger}={\cal S}{\cal S}_{}^{\dagger}+{\cal U}{\cal U}_{}^{\dagger}={\bf 1}$ and ${\cal V} {\cal S}_{}^{\dagger}+{\cal R} {\cal U}_{}^{\dagger}={\cal S} {\cal V}_{}^{\dagger}+{\cal U} {\cal R}_{}^{\dagger}={\bf 0}$. Note that in the limit of $M_i^{}\to \infty$, we have ${\cal R} \approx M^{}_{\rm D} M^{-1}_i \to 0$ and thus ${\cal V}$ is reduced to the unitary PMNS matrix~\cite{Pontecorvo1957,Maki1962}. The neutrino mass eigenstates $\left(\widehat{\nu}_{\rm L}^{}, \widehat{N}_{\rm R}^{\rm C}\right)$ and the flavor eigenstates $\left(\nu_L^{},N_{\rm R}^{\rm C}\right)$ are related by
\begin{eqnarray}
\left(
\begin{matrix}
\nu_{\rm L}^{}\\
N_{\rm R}^{\rm C}
\end{matrix}
\right)=
\left(
\begin{matrix}
{\cal V}& {\cal R}\\
{\cal S}& {\cal U}
\end{matrix}
\right)
\left(
\begin{matrix}
\widehat{\nu}_{\rm L}^{}\\
\widehat{N}_{\rm R}^{\rm C}
\end{matrix}
\right)\;,
\label{eq:seesawmixing}
%     (48)
\end{eqnarray}
and the interaction term relevant to the muon anomalous magnetic moment is given by
\begin{eqnarray}
{\cal L}_{\rm int}^{\rm Type-I}=\frac{g}{2\sqrt{2}}\overline{\mu}\gamma_{}^{\mu}\left(1-\gamma_5^{}\right)\left({\cal V}_{\mu i}^{} \widehat{\nu}_{i}^{} + {\cal R}_{\mu i}^{} \widehat{N}_i^{}\right) W_{\mu}^{-}+{\rm h.c.}\;,
%     (49)
\label{eq:seesawCC}
\end{eqnarray}
where ${\cal V}_{\alpha i}^{}$ and ${\cal R}_{\alpha i}^{}$ denotes the $(\alpha,i)$-element of ${\cal V}$ and ${\cal R}$ (for $\alpha=e,\mu,\tau$ and $i=1,2,3$). As indicated by Eq.~(\ref{eq:seesawCC}), due to the light-heavy neutrino mixing in Eq.~(\ref{eq:seesawmixing}), both light and heavy Majorana neutrinos are present in the charged-current interaction.

The contributions to the muon anomalous magnetic moment in the type-I seesaw model are of the FV-type and depicted in Fig.~\ref{fig:Type-I seesaw}. For the heavy neutrinos, using Eq.~(\ref{eq:gFV2}) with the identification of $Q_V^{}=-1$, $M_V^{}=M_W^{}$, $M_F^{}=M_i^{}$ and $C_{\rm V}^{}=-C_{\rm A}^{}=g{\cal R}_{\mu i}^{*}/(2\sqrt{2})$, we obtain
\begin{eqnarray}
a_{\mu}^{N}=\frac{G_{\rm F}^{}m_{\mu}^2}{4\sqrt{2}\pi^2}\sum_{i=1}^3 \left|{\cal R}_{\mu i}\right|_{}^2 {\cal G}\left(\frac{M_i^2}{M_W^2}\right)\;,
%     (50)
\end{eqnarray}
where the loop function ${\cal G}(x)$ comes from the second-order loop function in Eq.~(\ref{eq:gFV2}), i.e.,
\begin{eqnarray}
{\cal G}(x)=\frac{1}{2}g_{\rm FV}^{(2)}\left(\sqrt{x},-1\right)=\frac{10-43x+78x_{}^2-49x_{}^3+4x_{}^4+18x_{}^3\ln x}{6 \left(1-x\right)^4}\;,
%     (51)
\end{eqnarray}
which is always positive and monotonically decreasing with the limits ${\cal G}(0)=5/3$ and ${\cal G}(\infty)=2/3$. Thus the pure contributions from the heavy neutrinos to $a_\mu^{}$ are positive, as a consequence of the previous observation that $g_{\rm FV}^{(2)}(y,Q_V^{}) > 0$ if the electric charge of $W_{}^{-}$ is smaller than the lower critical charge, i.e., $Q^{}_{W^-} = -1 < Q^{\rm L}_V = - 5/9$ (cf. Table~\ref{table:FV}).

For the light neutrinos, making use of Eq.~(\ref{eq:FV appro2}) with $Q_V^{}=-1$, $M_V^{}=M_W^{}$, $M_F^{}=m_i^{}$ and $C_{\rm V}^{}=-C_{\rm A}^{}=g{\cal V}_{\mu i}^{*}/(2\sqrt{2})$, one gets
\begin{eqnarray}
\label{eq:Type-I seesaw light neutrino}
a_{\mu}^{\nu}=\frac{G_{\rm F}^{}m_{\mu}^2}{4\sqrt{2}\pi^2}\times\frac{5}{3}\sum_{i=1}^3 \left|{\cal V}_{\mu i}\right|_{}^2\;.
%     (52)
\end{eqnarray}
The contributions from the SM neutrinos can be obtained from Eq.~(\ref{eq:Type-I seesaw light neutrino}) simply by imposing the unitarity condition $\left|{\cal V}_{\mu 1}\right|_{}^2 + \left|{\cal V}_{\mu 2}\right|_{}^2 + \left|{\cal V}_{\mu 3}\right|_{}^2 = 1$, which is valid in the decoupling limit of $M_i\rightarrow \infty$. As a result, we have
\begin{eqnarray}
a_{\mu}^{\nu,{\rm SM}}=\frac{G_{\rm F}^{}m_{\mu}^2}{4\sqrt{2}\pi^2}\times\frac{5}{3}\;.
%     (53)
\end{eqnarray}
The final result for the extra contribution to muon $g-2$ in the type-I seesaw model can be obtained by subtracting the SM neutrino contribution, namely,
\begin{eqnarray}
\label{eq:Type-I seesaw}
\Delta a_{\mu}^{\rm Type-I}=a_{\mu}^N+a_{\mu}^{\nu}-a_{\mu}^{\nu,{\rm SM}}=\frac{G_{\rm F}^{}m_{\mu}^2}{4\sqrt{2}\pi^2}\sum_{i=1}^3\left\{
\left|{\cal R}_{\mu i}^{}\right|_{}^2\left[{\cal G}\left(\frac{M_i^2}{M_W^2}\right)-\frac{5}{3}\right]\right\}\;,
%     (54)
\end{eqnarray}
which is obviously negative because of the upper bound on the loop function (i.e., ${\cal G}(x) < 5/3$ for $x > 0$). Note that the  unitarity condition ${\cal V} {\cal V}_{}^{\dagger}+{\cal R}{\cal R}_{}^{\dagger}={\bf 1}$ has been used. The result in Eq.~(\ref{eq:Type-I seesaw}) is the same as that derived in Ref.~\cite{Zhou2021}, where the method of Feynman parametrization has been implemented instead.

\subsection{Type-II seesaw model}
\label{subsec:type-II seesaw}
%%%%%%%%%%%%%%%%%%%%%%%%%%%%%%% Fig. 8 %%%%%%%%%%%%%%%%%%%%%%%%%%%%%%%%
\begin{figure}[t!]
	\centering		\includegraphics[width=20cm]{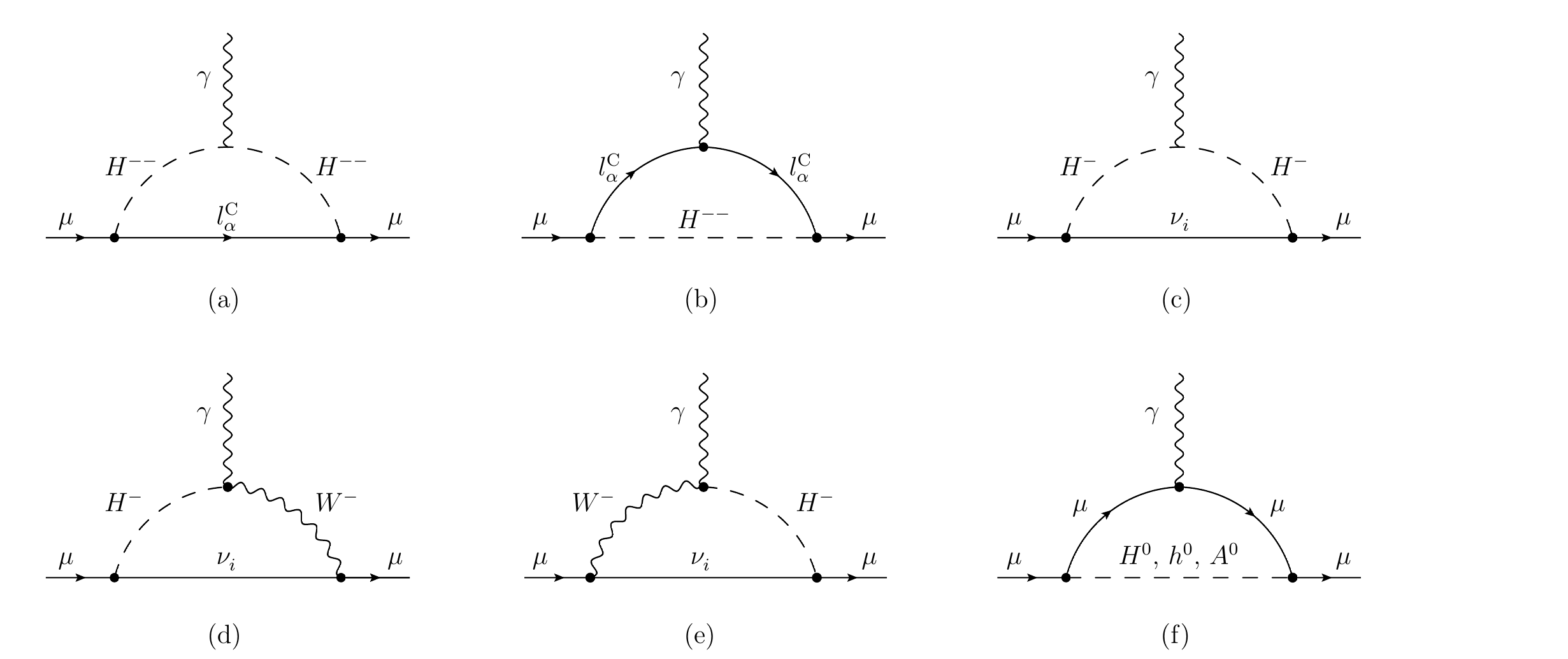}
		\vspace{-0.3cm}
	\caption{The Feynman diagrams for the contributions from the charged and neutral scalars to the muon anomalous magnetic moment in the type-II seesaw model: (a) and (b) from the doubly-charged scalar $H_{}^{--}$ and the charged leptons $l_{\alpha}^{\rm C}$ (for $\alpha=e,\mu,\tau$); (c), (d) and (e) from the singly-charged scalar $H_{}^{-}$ and the neutrinos $\nu_i^{}$ (for $i=1,2,3$); (f) from the neutral scalars, where $H_{}^0$, $h_{}^0$ and $A_{}^0$ are two CP-even and one CP-odd neutral scalars coming from the mixing between the SM Higgs boson and the neutral components of the triplet scalar.}
	\label{fig:Type-II seesaw}
\end{figure}
Another intriguing way to generate tiny neutrino masses is the type-II seesaw model~\cite{Konetschny1977,Magg1980,Schechter1980,Cheng1980,Mohapatra1981,Lazarides1981}, which augments the SM with a triplet scalar with the hypercharge $-1$, i.e.,
\begin{eqnarray}
\Delta=\sqrt{2}
\left(
\begin{matrix}
\xi_{}^{-}/\sqrt{2}& -\xi_{}^{0}\\
\xi_{}^{--}&-\xi_{}^{-}/\sqrt{2}
\end{matrix}
\right)\;.
%     (55)
\end{eqnarray}
The triplet has been written in the adjoint representation and the electric charge of each component has been explicitly indicated.

The neutral components of the triplet (i.e., $\xi_{}^0$ and its complex conjugate) mix with those of the SM Higgs doublet, leading to two CP-even scalar bosons $(H^0, h^0)$ and one CP-odd scalar boson $A^0$. The Feynman diagrams for their contributions to $a_\mu^{}$ are shown in Fig.~\ref{fig:Type-II seesaw}(f), which however are highly suppressed by the Yukawa coupling of muon. The leading contributions are made by the singly- and doubly-charged scalars and the relevant interaction reads
\begin{eqnarray}
{\cal L}_{\rm int}^{\rm Type-II}&=&\left[-\frac{1}{2}\cos\beta\left(Y_{\Delta}^{\dagger}
{\cal V}\right)_{\mu i}\overline{\hat{\nu}_i^{}}\left(1-\gamma_5^{}\right)\mu H_{}^{+}-\frac{1}{\sqrt{2}}\left(Y_{\Delta}\right)_{\alpha\mu}^{*}\overline{l_{\alpha}^{\rm C}}\left(1-\gamma_5^{}\right)\mu H_{}^{++}\right.\nonumber\\
&&\left. \frac{g}{2\sqrt{2}}{\cal V}_{\mu i}^{*}\overline{\nu_i^{}}\gamma_{}^{\mu}\left(1-\gamma_5^{}\right)\mu W_{\mu}^{+}+{\rm h.c.}\right] -\sqrt{2}g e v_{\Delta}^{}\cos\beta\left(H_{}^{-}W_{\mu}^{+}+H_{}^{+}W_{\mu}^{-}\right)A_{}^{\mu}
\;, \quad
%     (56)
\end{eqnarray}
where $Y_{\Delta}^{}$ represents the Yukawa coupling matrix between the scalar triplet and the lepton doublet, and ${\cal V}$ is the unitary PMNS matrix. The flavor eigenstates $\left(\xi_{}^{\pm}, \xi_{}^{\pm\pm}\right)$ are related to the mass eigenstates $\left(H_{}^{\pm},H_{}^{\pm\pm}\right)$ via
\begin{eqnarray}
\xi_{}^{\pm}=H_{}^{\pm}\cos\beta- G_{}^{\pm}\sin\beta\;,\qquad
\xi_{}^{\pm\pm}=H_{}^{\pm\pm}\;,
%     (57)
\end{eqnarray}
with $\beta=\cos_{}^{-1}\left(v/\sqrt{v_{}^2+2v_{\Delta}^2} \right)$ and $v_{\Delta}^{}$ being the VEV of $\Delta$. The Feynman diagrams contributing to $a^{}_\mu$ are shown in Fig.~{\ref{fig:Type-II seesaw}}. The diagrams involving doubly-charged scalars can be found in Fig.~{\ref{fig:Type-II seesaw}}(a) and Fig.~{\ref{fig:Type-II seesaw}}(b). Noticing $m_{\alpha}^{}\ll M_{H^{\pm\pm}}^{}$ (for $\alpha=e,\mu,\tau$) and utilizing Eq.~(\ref{eq:FS appro2}) with $C_{\rm S}^{}=-C_{\rm P}^{}=-\left(Y_{\Delta}^{}\right)_{\alpha\mu}^{*}/\sqrt{2}$, $Q_S^{}=-2$, $M_S^{}=M_{H^{\pm\pm}}^{}$ and $M_F^{}=m_{\alpha}^{}$, one can immediately obtain
\begin{eqnarray}
\Delta a_{\mu}^{H^{\pm\pm}}=-\frac{1}{12\pi^2}\left(Y_{\Delta}^{\dagger}Y_{\Delta}^{}\right)_{\mu\mu}\left(\frac{m_{\mu}}{M_{H^{\pm\pm}}}\right)_{}^2\;.
\label{eq:typeII-Hpp}
%     (58)
\end{eqnarray}
In addition, Figs.~{\ref{fig:Type-II seesaw}}(c), (d) and (e) represent the contributions from the singly-charged scalars. For Fig.~{\ref{fig:Type-II seesaw}}(c), we can also use Eq.~(\ref{eq:FS appro2}) but with $C_{\rm S}^{}=-C_{\rm P}^{}=-\cos\beta\left(Y_{\Delta}^{\dagger}{\cal V}\right)_{\mu i}^{}/2 $, $Q_S^{}=-1$, $M_S^{}=M_{H^{\pm}}^{}$ and $M_F^{}=m_{\alpha}^{}$ to derive
\begin{eqnarray}
\label{eq:Type-II c}
\Delta a_{\mu}^{H^{\pm}({\rm c})}=-\frac{1}{96\pi^2}\left(Y_{\Delta}^{\dagger}Y_{\Delta}^{}\right)_{\mu\mu}\left(\frac{m_{\mu}}{M_{H^{\pm}}}\right)_{}^2\;,
%     (59)
\end{eqnarray}
where we have used the experimental bound $v_{\Delta}^{}\ll v$ coming from the precision measurements of the $\rho$ parameter. For Fig.~{\ref{fig:Type-II seesaw}}(d) and Fig.~{\ref{fig:Type-II seesaw}}(e), which belong to the FSV-type diagrams, we notice that $m_{\mu}^{}\ll M_{H^{\pm}}^{},M_W^{}$ is perfectly satisfied and then make use of Eq.~(\ref{eq:FSV appro}) with $C_{\rm S}^{}=-C_{\rm P}^{}=-\cos\beta\left(Y_{\Delta}^{\dagger}{\cal V}\right)_{\mu i}^{}/2$, $C_{\rm V}^{}=-C_{\rm A}^{}=g {\cal V}_{\mu i}^{*}/\left(2\sqrt{2}\right)$, $\mu_0^{}=-\sqrt{2}gv_{\Delta}^{}\cos\beta$, $M_S^{}=M_{H^{\pm}}^{}$, $M_V^{}=M_{W}^{}$ and $M_F^{}=m_{\alpha}^{}$. The final result is
\begin{eqnarray}
\label{eq:Type-II de}
\Delta a_{\mu}^{H^{\pm }(\rm{d})+(\rm{e})} = \frac{G_{\rm F}v_{\Delta}^2}{6\sqrt{2}\pi^2} \frac{m^2_\mu \ln\left(M^2_W/M^2_{H^\pm}\right) }{M^2_W - M^2_{H^\pm}} \left(Y_{\Delta}^{\dagger}Y_{\Delta}^{}\right)_{\mu\mu}\;,
%     (60)
\end{eqnarray}
where is highly suppressed by the tiny neutrino masses due to $G^{}_{\rm F} v^2_\Delta \propto m^2_i/M^2_W$. Therefore, the contribution in Eq.~(\ref{eq:Type-II de}) can be safely neglected, and the overall result is given by
\begin{eqnarray}
\Delta a_{\mu}^{H^{\pm}}\approx \Delta a_{\mu}^{H^{\pm}({\rm c})}=-\frac{1}{96\pi^2}\left(Y_{\Delta}^{\dagger}Y_{\Delta}^{}\right)_{\mu\mu}\left(\frac{m_{\mu}}{M_{H^{\pm}}}\right)_{}^2\;.
\label{eq:typeII-Hp}
%     (61)
\end{eqnarray}

As shown in Eqs.~(\ref{eq:typeII-Hpp}) and (\ref{eq:typeII-Hp}), the dominant contributions to the muon anomalous magnetic moment in the type-II seesaw turn out to be negative in the whole parameter space of $Y_{\Delta}$. The same results have also been obtained in Ref.~\cite{Fukuyama2009}. It is worthwhile to mention that this wrong sign can be simply observed from the second-order loop function in Eq.~(\ref{eq:gFS2}). Since the electric charges of the charged scalars in the type-II seesaw model are smaller than its lower critical charge $Q_{S}^{\rm L}=-2/3$ (cf. Table~\ref{table:FS}), we have
\begin{eqnarray}
g_{\rm FS}^{(2)}\left(y,Q_S^{}=-1\right)<0\;,\qquad
g_{\rm FS}^{(2)}\left(y,Q_S^{}=-2\right)<0\;,
\end{eqnarray}
implying the overall negative sign of $\Delta a^{}_\mu$. Note that $|C^{}_{\rm S}|^2 - |C^{}_{\rm P}|^2 = 0$ holds for both doubly- and singly-charged scalars, so the first-order loop function in Eq.~(\ref{eq:gFS1}) is irrelevant.

%%%%%%%%%%%%%%%%%%%%%%%%%%%%%%%%%%%%%%%%%%%%%%%%%%%%%%%%%%%%%%%%%%%%%%%
\subsection{$Z_{}^{\prime}$ model}
%%%%%%%%%%%%%%%%%%%%%%%%%%%%%%% Fig. 9 %%%%%%%%%%%%%%%%%%%%%%%%%%%%%%%%
\begin{figure}[t!]
	\centering		\includegraphics[width=9cm]{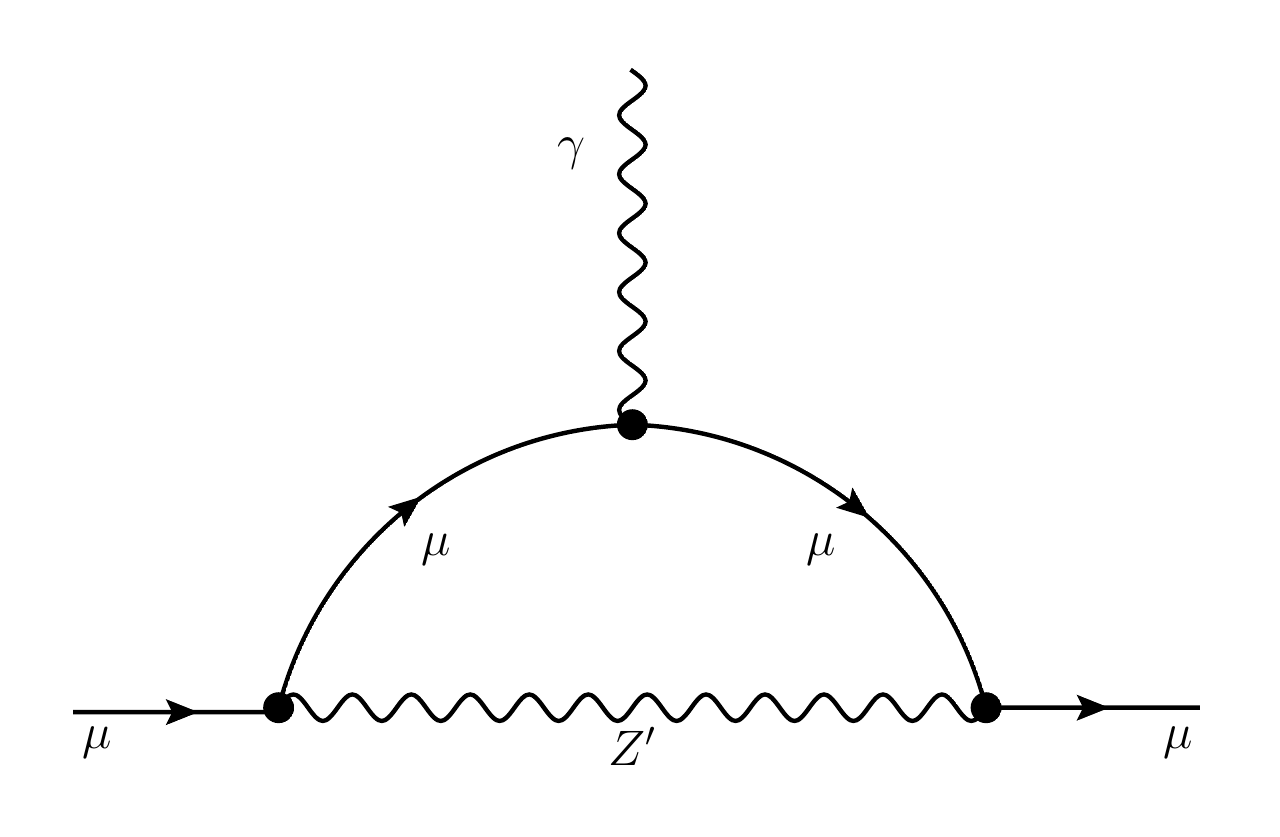}
		\vspace{-0.3cm}
	\caption{The Feynman diagram for the contribution from the new gauge boson $Z_{}^{\prime}$ to the muon anomalous magnetic moment.}
	\label{fig:Z prime}
\end{figure}
%%%%%%%%%%%%%%%%%%%%%%%%%%%%%%%%%%%%%%%%%%%%%%%%%%%%%%%%%%%%%%%%%%%%%%%
The linear combinations $B/3-L_{\alpha}^{}$, $L_e^{}-L_{\mu}$ and $L_\mu^{}-L_\tau^{}$ of the baryon number $B$ and the lepton number $L_{\alpha}^{}$ (for $\alpha=e,\mu,\tau$) are anomaly-free, so any one of them can naturally be promoted to a gauge symmetry\footnote{Note that $L_e^{}-L_{\mu}$ and $L_\mu^{}-L_\tau^{}$ can be gauged without including right-handed neutrinos, whereas $B/3-L_{\alpha}^{}$ (for $\alpha=e,\mu,\tau$) can be gauged only when at least one right-handed neutrino is introduced.}~\cite{Foot1991,Foot1991p,He1991,He1991p}. The neutral gauge boson $Z_{}^{\prime}$ corresponding to the new ${\rm U}(1)$ gauge symmetry mediates extra interaction among charged leptons and can contribute to the muon $g-2$. For illustration, we assume that the coupling is flavor-diagonal and both $\left(\nu_{\mu}^{},\mu \right)_{\rm L}^{}$ and $\mu_{\rm R}^{}$ take the same charge under this new ${\rm U}(1)$ gauge group. The relevant Lagrangian is
\begin{eqnarray}
{\cal L}_{\rm int}^{Z^{\prime}}=g_{Z^{\prime}}^{}\overline{\mu}\gamma_{}^{\mu}\mu Z_{\mu}^{\prime}\;,
%     (63)
\end{eqnarray}
with $g_{Z^{\prime}}^{}$ the coupling constant. The Feynman diagram contributing to the muon $g-2$ is shown in Fig.~\ref{fig:Z prime}. By inserting $C_{\rm V}^{}=g_{Z^{\prime}}^{}$, $C_{\rm A}^{}=0$, $Q_V^{}=0$, $M_V^{}=M_{Z^{\prime}}^{}$ and $M_F^{}=m_{\mu}^{}$ into Eqs.~(\ref{eq:FV amu})-(\ref{eq:FV+}), one obtains
\begin{eqnarray}
\label{eq:Z prime}
\Delta a_{\mu}^{Z^{\prime}}=\frac{g_{Z^{\prime}}^2}{16\pi^2} {\cal H}\left(\frac{m_{\mu}}{M^{}_{Z^{\prime}}}\right)\;,
%     (64)
\end{eqnarray}
where the loop function
\begin{eqnarray}
\label{eq:Z prime loop function}
{\cal H}(x)=\frac{2}{x^4}\left[x_{}^2\left(x_{}^2-2\right)+2\left(2x_{}^2-1\right)\ln x+2x_{}^2\left(4x_{}^2-2x_{}^4-1\right)\frac{\Lambda\left(x^2,x,1\right)}{\lambda\left(x^2,x^2,1\right)}
\right]\;.
%    (65)
\end{eqnarray}
It is easy to verify that the loop function in Eq.~(\ref{eq:Z prime loop function}) is non-negative with the following finite limits
\begin{eqnarray*}
\lim_{x\to 0}^{}{\cal H}(x)=0,\quad \lim_{x\to \infty}^{}{\cal H}(x)=2 \;.
\end{eqnarray*}
Thus one can reach the conclusion that the neutral vector boson with a vector-like interaction with muon will always make a positive contribution to $a_{\mu}^{}$. The formulae in Eqs.~(\ref{eq:Z prime}) and (\ref{eq:Z prime loop function}) are valid for the whole parameter space of $M^{}_{Z^{\prime}}$. In particular, in the hierarchical cases of $m_{\mu}^{}\ll M^{}_{Z^{\prime}}$ and $M^{}_{Z^{\prime}} \ll m_{\mu}^{}$, we have
\begin{eqnarray}
	\label{eq:Z prime hierarchy}
\Delta a_{\mu}^{Z^{\prime}}=
\left\{
  \begin{array}{ll}
\displaystyle \frac{g_{Z^{\prime}}^2}{12\pi^2}\left(\frac{m_{\mu}}{M_{Z^{\prime}}}\right)_{}^2 \;, & \hbox{$m^{}_\mu \ll M^{}_{Z^\prime}$;} \\
~ & ~ \\
\displaystyle \frac{g_{Z^{\prime}}^2}{8\pi^2}\;, & \hbox{$M^{}_{Z^\prime} \ll m^{}_\mu$.}
  \end{array}
\right.
%     (65)
\end{eqnarray}
Moreover, in the limit of $M^{}_{Z^\prime} \to m^{}_\mu$ or equivalently $x = m^{}_\mu/M^{}_{Z^\prime} \to 1$, we can obtain ${\cal H}(x)|^{}_{x \to 1} = 2(2\pi - 3\sqrt{3})/(3\sqrt{3})$ from Eq.~(\ref{eq:Z prime loop function}). Note that the result in Eq.~(\ref{eq:Z prime hierarchy}), which has previously been obtained in the literature~\cite{Zhou2021, Bodas2021}, is valid only in the assumption of a strong mass hierarchy. For a complete scan of the parameter space of $M_{Z_{}^\prime}^{}$ in the most general case, it is necessary to use the exact formulae in Eqs.~(\ref{eq:Z prime})-(\ref{eq:Z prime loop function}).

\subsection{Leptoquark}
%%%%%%%%%%%%%%%%%%%%%%%%%%%%%%% Fig. 10 %%%%%%%%%%%%%%%%%%%%%%%%%%%%%%%%
\begin{figure}[t!]
	\centering		\includegraphics[width=17.5cm]{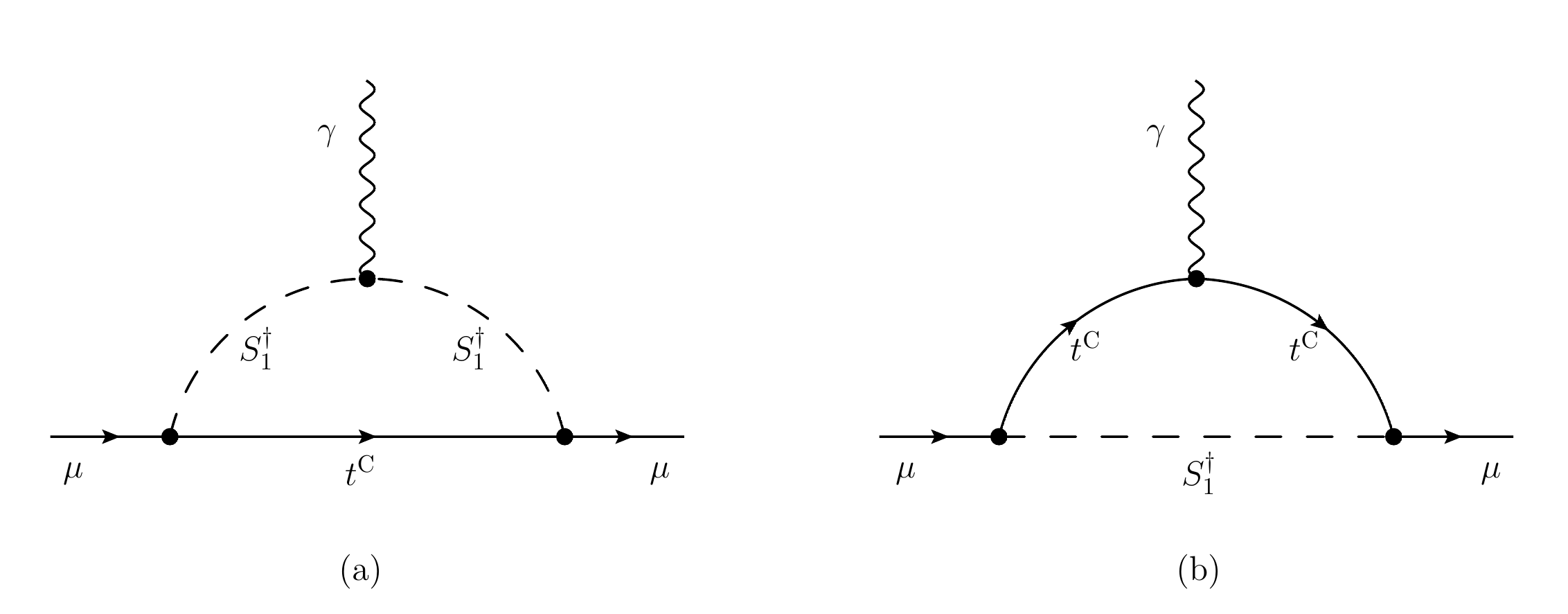}
		\vspace{-0.3cm}
	\caption{The Feynman diagrams for the contribution from the scalar leptoquark $S_1^{}$ to the muon anomalous magnetic moment.}
	\label{fig:leptoquark}
\end{figure}
%%%%%%%%%%%%%%%%%%%%%%%%%%%%%%%%%%%%%%%%%%%%%%%%%%%%%%%%%%%%%%%%%%%%%%%
Leptoquarks are hypothetical scalar or vector bosons that are coupled simultaneously to leptons and quarks. In the first place, leptoquarks arise naturally from the theories of grand unification. At present, they have been extensively discussed in the literature as new-physics explanations for, e.g., tiny neutrino masses via radiative corrections, $B$ anomalies and muon anomalous magnetic moment. See, for example, Ref.~\cite{Dorsner2016}, for a recent review on leptoquarks. For illustration, we consider the scalar leptoquark $S_1^{}$, whose quantum numbers are assigned as $\left({\bf \overline{3}}, {\bf 1}, 1/3\right)$ under the ${\rm SU}(3)_{\rm C}^{}\otimes {\rm SU}(2)_{\rm L}^{}\otimes {\rm U}(1)_{\rm Y}^{}$ gauge group. The main contribution to the muon anomalous magnetic moment comes from the interaction among muon, leptoquark and top quark, as shown in Fig.~\ref{fig:leptoquark}. The relevant Lagrangian can be written as
\begin{eqnarray}
{\cal L}_{\rm int}^{S_1^{}}=\overline{t_{}^{\rm C}}\left(C_{\rm S}^{}+C_{\rm P}^{}\gamma_5^{}\right)\mu S_1^{}+ {\rm h.c.}\;.
%     (67)
\end{eqnarray}

Assuming $S_1^{}$ to be much heavier than muon and substituting $Q_S^{}=Q_{S_1^{\dagger}}^{}=-1/3$, $M_S^{}=M_{S_1}^{}$ and $M_F^{}=m_t^{}$ into Eq.~(\ref{eq:gFS1}), we arrive at
\begin{eqnarray}
\Delta a_{\mu}^{S_1^{}}=\frac{1}{16\pi^2}\left(\left|C_{\rm S}\right|_{}^2-\left|C_{\rm P}^{}\right|_{}^2\right)\frac{m_{\mu}}{M_{S_1}}{\cal I}\left(\frac{m_t^{}}{M_{S_1}}\right)\;,
\label{eq:DeltaamuS1}
%     (68)
\end{eqnarray}
where the loop function ${\cal I}(x)$ is derived from the first-order loop function in Eq.~(\ref{eq:gFS1}), i.e.,
\begin{eqnarray}
{\cal I}\left(x\right)=3g_{\rm FS}^{(1)}\left(x,-\frac{1}{3}\right)=\frac{x\left[x^4-8x^2+7+4\left(x^2+2\right)\ln x \right]}{\left(x^2-1\right)^3}\;,
%     (69)
\label{eq:DeltaamuS1loop}
\end{eqnarray}
with the color factor $3$ being included. Notice that we have assumed $S_1^{}$ to be coupled to both left- and right-handed muons, implying that the contribution to $a^{}_\mu$ is dominated by the first-order loop function. Two helpful comments are in order. First, the sign of $\Delta a_{\mu}^{S_1^{}}$ in Eq.~(\ref{eq:DeltaamuS1}) depends on the relative sizes of $\left|C_{\rm S}^{}\right|$ and $\left|C_{\rm P}^{}\right|$, since the function $g_{\rm FS}^{(1)}(x, -1/3)$ is always positive. The latter can be understood by noticing that the electric charge of $S_1^{\dagger}$ is larger than the upper critical charge, i.e., $Q_{S_1^{\dagger}}^{} = -1/3 > Q_S^{\rm U} = -1/2$ (cf. Table~\ref{table:FS}). Second, if we further assume that $S_1^{}$ is also much heavier than top quark, then the expression in Eq.~(\ref{eq:DeltaamuS1}) will be reduced to
\begin{eqnarray}
\Delta a_{\mu}^{S_1^{}}=\frac{1}{16\pi^2}\left(\left|C_{\rm S}\right|_{}^2-\left|C_{\rm P}^{}\right|_{}^2\right)\frac{m_{\mu}m_t}{M_{S_1}^2}\left[-7-8\ln\left(\frac{m_t}{M_{S_1}}\right) \right]\;.
%     (70)
\end{eqnarray}
This result has previously been derived in the literature ~\cite{ColuccioLeskow2016,Crivellin2020,Marzocca2021}. However, for a complete scan of the parameter space, it is necessary to utilize the general expression of the loop function in Eq.~(\ref{eq:DeltaamuS1loop}).

\section{Summary}
\label{sec:summary}
Motivated by the recent measurement of the muon anomalous magnetic moment $a^{}_\mu$, which is at odds with the SM prediction at the $4.2~\sigma$ level, we examine possible new-physics contributions to $a^{}_\mu$ at the one-loop order. A model-independent analysis of the general loop functions has been carried out. The main results in the present paper are summarized as follows.

First, the contributions to $a^{}_\mu$ at the one-loop order can be classified into three categories by different types of particles running in the loop: (1) The FS-type with one fermion and one scalar boson; (2) The FV-type with one fermion and one vector boson; (3) The FSV-type with one fermion, one scalar boson and one vector boson. Although the one-loop Feynman diagrams for $a^{}_\mu$ have been studied for a long time, it is surprising that the general formulae have never been explicitly given in the literature. Instead of expressing the final results in terms of the integral over the Feynman parameter, we use the Passarino-Veltman functions to derive a general and compact formula for $a_\mu^{}$ without any integrals in the end. The final results can be found in Eqs.~(\ref{eq:FS amu})-(\ref{eq:FS+}) for the FS-type, Eqs.~(\ref{eq:FV amu})-(\ref{eq:FV+}) for the FV-type, and Eqs.~(\ref{eq:FSV amu})-(\ref{eq:FSV+}) for the FSV-type. 

Second, we attempt to clarify when the new-physics contribution to $a^{}_\mu$ is positive so as to reconcile the theoretical prediction with the experimental measurement. For this purpose, we have investigated the basic properties of the loop functions of the FS- and FV-type. In the assumption that the new particles are much heavier than muon, the loop functions can be series expanded in terms of the mass ratio $m^{}_\mu/M^{}_S$ (for the FS-type) or $m^{}_\mu/M^{}_V$ (for the FV-type). At different orders of $m^{}_\mu/M^{}_S$ or $m^{}_\mu/M^{}_V$, the loop functions exhibit interesting behaviors depending on the electric charges of the particles running in the loop. Roughly speaking, there exists an upper and a lower critical value for the electric charge $Q^{}_S$ or $Q^{}_V$, which can be used to claim whether the contribution to $a^{}_\mu$ is positive or negative. The properties of the first- and second-order loop functions are summarized in Table~\ref{table:FS} and Table~\ref{table:FV}. These properties have also been applied to make a quick judgment about the sign of $\Delta a_{\mu}^{}$ in a concrete model.

Third, it is interesting to find that there are two extra constraints for the FV-type model. First, if there exists in the theory a heavy fermion that is coupled simultaneously to both left- and right-handed muon and the vector boson, then the couplings should decrease no more slowly than ${\cal O}\left(1/\sqrt{M_F}^{}\right)$ in the limit $M^{}_F \to \infty$. Second, according to the Weinberg-Witten theorem, there cannot be a massless vector boson with a nonvanishing electric charge, contributing to muon anomalous magnetic moment. These constraints may be automatically satisfied in a self-consistent theory, but should be carefully taken into account in phenomenological studies.

Finally, our general formulae have been applied to several well-motivated theoretical models. The correct results in all these models can be reproduced in a straightforward way. If the $4.2~\sigma$ discrepancy between the theoretical prediction and the experimental measurement of $a^{}_\mu$ is further confirmed, new physics will definitely be introduced to explain such a discrepancy. In this case, the main results presented in our paper will be helpful for both model building and phenomenological studies associated with the muon anomalous magnetic moment.

\section*{Acknowledgements}

This work was supported in part by the National Natural Science Foundation of China under grant No.~11775232 and No.~11835013, by the Key Research Program of the Chinese Academy of Sciences under grant No. XDPB15, and by the CAS Center for Excellence in Particle Physics.

\end{document}